\documentclass[usenatbib,letterpaper]{mn2e}
\usepackage{amsmath}
\usepackage{amssymb}
\usepackage{amsbsy}
\usepackage{graphicx}
\usepackage[colorlinks,citecolor=blue]{hyperref}
\usepackage{ifthen}  
\usepackage{tensor}
\usepackage{rotating}
\usepackage{xcolor}
\usepackage{footnote}
\usepackage{calc}
\usepackage{multicol}
\usepackage{color}
\usepackage{colortbl}

\newcounter{loRes}
\newcounter{extra}
\def\glc{{\sc Galacticus}}

\newcounter{CDMDone}
\def\CDM{\ifthenelse{\equal{\arabic{CDMDone}}{0}}{cold dark matter (CDM)\setcounter{CDMDone}{1}}{CDM}}

\newcounter{MCMCDone}
\def\MCMC{\ifthenelse{\equal{\arabic{MCMCDone}}{0}}{Markov Chain Monte Carlo (MCMC)\setcounter{MCMCDone}{1}}{MCMC}}

\newcounter{PPCDone}
\def\PPC{\ifthenelse{\equal{\arabic{PPCDone}}{0}}{posterior predictive check (PPC)\setcounter{PPCDone}{1}}{PPC}}
\def\PPCs{\ifthenelse{\equal{\arabic{PPCDone}}{0}}{posterior predictive checks (PPCs)\setcounter{PPCDone}{1}}{PPCs}}

\newcounter{PPDDone}
\def\PPD{\ifthenelse{\equal{\arabic{PPDDone}}{0}}{posterior probability distribution (PPD)\setcounter{PPDDone}{1}}{PPD}}

\newcounter{SAMDone}
\def\SAM{\ifthenelse{\equal{\arabic{SAMDone}}{0}}{semi-analytic model (SAM)\setcounter{SAMDone}{1}}{SAM}}
\def\SAMs{\ifthenelse{\equal{\arabic{SAMDone}}{0}}{semi-analytic models (SAMs)\setcounter{SAMDone}{1}}{SAMs}}

\newcounter{MDPLDone}
\def\MDPL{\ifthenelse{\equal{\arabic{MDPLDone}}{0}}{MultiDark Planck N-body (MDPL)\setcounter{MDPLDone}{1}}{MDPL}}

\title[]{The Mass Function of Unprocessed Dark Matter Halos and Merger Tree Branching Rates}
\author[Andrew J. Benson]{Andrew J. Benson$^1$\\
$^1$ Carnegie Observatories, 813 Santa Barbara Street, Pasadena, CA 91101, USA.}

\begin{document}

\maketitle

\begin{abstract}
A common approach in semi-analytic modeling of galaxy formation is to construct Monte Carlo realizations of merger histories of dark matter halos whose masses are sampled from a halo mass function. Both the mass function itself, and the merger rates used to construct merging histories are calibrated to N-body simulations. Typically, ``backsplash'' halos (those which were once subhalos within a larger halo, but which have since moved outside of the halo) are counted in both the halo mass function, and in the merger rates (or, equivalently, progenitor mass functions). This leads to a double-counting of mass in Monte Carlo merger histories which will bias results relative to N-body results. We measure halo mass functions and merger rates with this double-counting removed in a large, cosmological N-body simulation with cosmological parameters consistent with current constraints. Furthermore, we account for the inherently noisy nature of N-body halo mass estimates when fitting functions to N-body data, and show that ignoring these errors leads to a significant systematic bias given the precision statistics available from state-of-the-art N-body cosmological simulations.
\end{abstract}


\section{Introduction}\label{sec:Introduction}

In the \CDM\ cosmogony, structure forms through the hierarchical merging of dark matter halos \citep{davis_evolution_1985}. The distribution of halo masses at a given epoch in the universe is a topic of significant interest and has been studied extensively \citep{press_formation_1974,bond_excursion_1991,jenkins_mass_2001,sheth_excursion_2002,white_mass_2002,lukic_halo_2007,reed_evolution_2003,warren_precision_2006,reed_halo_2007,tinker_toward_2008}. The standard definition of a halo is a region of dark matter which has undergone gravitational collapse and which exists in a state at least close to virial equilibrium. Practically, halos are typically defined as regions around density peaks whose mean density exceeds some threshold value, often motivated by the simple spherical top-hat collapse model (e.g. \citealt{press_formation_1974}). Recently, \cite{despali_universality_2015} have shown that this choice of density threshold should be preferred as it results in the most universal mass function when expressed in terms of the rescaled mass variable\footnote{While this is good motivation for choosing a particular definition of halo (it is at least motivated by a simple physical model and, as \protect\cite{despali_universality_2015} show, leads to the most universal mass function), it is highly idealized, and there is no notable physical distinction to this radius. It may be more robust therefore to characterize dark matter halos by some other quantity, such as the peak velocity in their rotation curve, and construct a $V_{\rm max}$ function instead. Better still, halos could be characterized by a two parameter function, specifying their abundance as a function of $V_{\rm max}$ and $r_{\rm s}$ (the scale radius in the \protect\cite{navarro_universal_1997} density profile). Together those two parameters specify the complete density profile of a halo, allowing its mass under any definition to be easily computed, and does not rely on a specific mass definition. We address this issue further in Benson (2017, in preparation).}, $\nu=\delta_{\rm c}/\sigma(M)$, where $\delta_{\rm c}$ is the critical linear theory overdensity for a halo to undergo gravitational collapse, and $\sigma(M)$ is the amplitude of fluctuations in spheres containing, on average, mass $M$ in the linear density field at the present day. 

Since halos grow hierarchically it was long suspected that the remnants of earlier generations of halos would survive for some time inside the larger halos into which they merge. This was confirmed by \cite{moore_dark_1999} and \cite{klypin_galaxies_1999}, and is now a well-established fact in the \CDM\ paradigm. Since many of these subhalos can survive for many dynamical times, and initially cross the virial radius of the halo with non-zero radial velocity, it should not be surprising that some subhalos are able to exit the virial radius once again as their orbit carries them toward apocentre\footnote{Some halos will not exit the halo of course, due to the effects of dynamical friction, the time evolution of the host halo potential, and possible tidal destruction.}. These ``backsplash'' halos have also been identified in N-body simulations \citep{moore_age-radius_2004,gill_evolution_2005,warnick_tidal_2008,ludlow_unorthodox_2009} with around 30\% of subhalos with orbital pericentres lying within the virial radius being found outside the virial radius at $z=0$ \citep{gill_evolution_2005}. The backsplash population may differ significantly in its properties due to the environmental effects of the host halos through which they have passed \citep{knebe_luminosities_2011}. It could be argued, therefore, that these backsplash halos should be excluded from measurements of the mass function and treated as a separate population.

Furthermore, in studies which employ merger trees generated using the extended Press-Schechter methodology (\citealt{lacey_merger_1993}; or derivatives thereof, e.g. \citealt{parkinson_generating_2008}) to follow the hierarchical growth of structure these backsplash halos can be double counted. The branching rates used to construct merger trees are typically calibrated to reproduce the conditional mass functions (that is, the distribution of progenitor halo masses at some earlier time, conditioned on those progenitors being part of a halo of given mass at some given later time). These conditional mass functions will inevitably (and correctly) include some halos which later become backsplash halos. The problem arises when ensembles of merger trees are averaged over (e.g. to construct galaxy mass functions). Here, the weight assigned to each merger tree is proportional to the halo mass function in order to give a fair sampling. But, under standard definitions, the halo mass function will include those same backsplash halos which were already included in conditional mass functions to which the merger tree branching rates were calibrated. As a result, the population of backsplash halos is double counted. This will bias estimates of any quantity computed by averaging over an ensemble of merger trees constructed in this way, and will manifest itself whenever comparing results based on merger trees extracted from N-body simulations, and those based on trees built using extended Press-Schechter-based approaches to mimic the N-body trees.

In this work, we construct and examine halo mass functions from which backsplash halos have been excluded. We provide a calibration of the fitting function of \cite{sheth_excursion_2002} which approximately reproduces these ``backsplashless'' mass functions and which can therefore be used in computing unbiased averages over ensembles of merger trees. Additionally, we provide a recalibration of merger tree branching rates using the fitting function of PCH, for updated cosmological parameters, with a significantly larger calibration set and defined consistently with our halo mass function (i.e. with $z=0$ backsplash halos excluded\footnote{Standard extended Press-Schechter-type approaches do not provide spatial information on subhalos, and so cannot make predictions about the population of backsplash halos. Such predictions could plausibly be made however if the technique was augmented through the inclusion of a dynamical model for subhalo orbits \protect\citep{taylor_dynamics_2001,benson_effects_2002,zentner_physics_2005}.}).

The remainder of this paper is organized as follows. In \S\ref{sec:method} we describe the N-body data employed as our calibration dataset, the construction of halo mass functions and conditional mass functions, and our treatment of the errors in N-body halo mass estimates. We present the results of our calibrations in \S\ref{sec:results}, and discuss their implications in \S\ref{sec:discussion}. Finally, in \S\ref{sec:conclusions} we give our conclusions.

\section{Method}\label{sec:method}

In this section we describe how we measure mass functions from the \MDPL\ simulation \citep{klypin_multidark_2016}, and how we construct model mass functions and conditional mass functions. We make extensive use of the analysis facilities of the \glc\ galaxy formation toolkit \citep{benson_g_2012} to perform these tasks.

\subsection{N-body Data}

For this work we utilize the \MDPL\ simulation \citep{klypin_multidark_2016}. This simulation has a box size of 1.48~Gpc, contains $3840^3$ particles each of mass $1.70\times10^9{\rm M}_\odot$, and adopts cosmological parameters consistent with measurements from the \emph{Planck} satellite \citep{planck_collaboration_planck_2014}, namely $(H_0,\Omega_\Lambda,\Omega_{\rm M},\Omega_{\rm b},n,\sigma_8)=(67.77~\hbox{km/s/Mpc},0.6929,0.3071,0.04820,0.96,0.8228)$. Halos were identified using the {\sc RockStar} algorithm \citep{behroozi_rockstar_2013}, and merger trees were constructed from these halos using the {\sc ConsistentTrees} algorithm \citep{behroozi_gravitationally_2013}. {\sc RockStar} provides several masses for each halo---in this work we exclusively make use of the ``virial mass'' supplied by {\sc RockStar} which is defined as the mass within a sphere of mean density given by the fitting formula of \cite{bryan_statistical_1998} to spherical top-hat collapse solutions (and is therefore consistent with the preferred mass of \citealt{despali_universality_2015}). The resulting merger trees contain 88 time snapshots between $z=0$ and $z=9.34$.

\subsection{Halo Mass Functions}

The standard mass function can be found from an N-body simulation by counting halos into bins logarithmically spaced in halo mass. For the backsplashless mass function we must first identify, and then exclude, all backsplash halos---that is, any halo which was identified as a subhalo in any previous timestep. As we have merger trees this is a straightforward exercise. \glc\ computes for each halo in a merger tree both the current and highest-ever ``hierarchy level''. Hierarchy level is defined to be 0 for an isolated halo (i.e. a halo which is not contained within any other halo), and $N+1$ for a halo contained within a halo of hierarchy level $N$. Thus, level-1 halos are subhalos, level-2 are sub-subhalos, etc. A backsplash halo has a current hierarchy level of 0, and a maximum hierarchy level ever reached greater than 0.

We find that, while this definition of backsplash halos has the virtue of being simple, it leads to implausible behaviour. In particular, many very high mass halos are identified as backsplash halos. This happens because, very early in the history of the halo, its primary progenitor (at the time of low mass) can be identified as a subhalo---this is then recorded in the maximum hierarchy level reached all the way to the final halo. This subhalo identification is likely spurious and reflects the difficulties of assigning which is the primary halo \citep{srisawat_sussing_2013}.

Therefore, we reset the maximum hierarchy level to zero if a halo is currently level 0 and its current mass exceeds its mass when it was last a subhalo by a factor $f_{\rm reset}$---that is, if $M_{\rm current} > f_{\rm reset} M_{\rm sub}$. We would typically not expect backsplash halos to grow significantly in mass, as they (by definition) do not dominate their local potential. We choose $f_{\rm reset}=2$ as our canonical value. Our results are somewhat sensitive to this choice. At the lowest halo masses studied in this work (corresponding to 300 particles), we find that backsplash halos make up 7\% of all halos for $f_{\rm reset}=2$. Increasing $f_{\rm reset}$ to $4$ increases this fraction to 9\%, while decreasing $f_{\rm reset}$ to 1 (which is not recomended as any upward fluctuation in halo mass would then trigger the maximum hierarchy level to be reset to zero, but is included here as an extreme case) results in a 4\% backsplash fraction (for 3000 particle halos these fractions become 1.09\%, 1.43\%, and 0.36\% respectively, so that even at higher particle number the choice of $f_{\rm reset}$ still makes a significant difference in the fraction of backsplash halos, although that fraction is smaller overall). Improvement in this aspect of backsplash halo identification will require the development of more robust algorithms for merger tree construction and the assignment of primary halo status.

To model the resulting backsplashless mass function we use the parametric form proposed by \cite{sheth_excursion_2002},
\begin{eqnarray}
{{\rm d}n\over {\rm d}M}(M) &=& {A \Omega_{\rm M} \rho_{\rm c} \over M^2} \left|{{\rm d}\log \sigma \over {\rm d}\log M} (M)\right|\left({2\over \pi} \nu^\prime\right)^{1/2} \left( 1 + {1 \over \nu^{\prime p}} \right) \nonumber \\
 & & \times \exp\left(-{1\over2}\nu^\prime\right),
\end{eqnarray}
with $\nu^\prime = a \nu^2$, but convolved with the N-body halo mass error distribution described in \S\ref{sec:SOerrors}, allowing the three parameters $a$, $p$, and $A$ (the overall normalization) to vary. This parametric form was also used by \cite{despali_universality_2015}. To find the optimal values of the parameters $(a,p,A)$ we perform a \MCMC\ simulation in which these parameters are allowed to vary.

\subsection{Merger Tree Branching Rates}

To constrain the branching rates in merger trees we first construct conditional halo mass functions from the \MDPL\ simulation. In particular, we measure from the \MDPL\ the distribution of progenitor halo masses at several redshifts $z>0$ of halos selected in mass bins at $z=0$. We then use the algorithm of \citeauthor{parkinson_generating_2008}~(\citeyear{parkinson_generating_2008}; hereafter PCH) to construct merger trees spanning the same range of halo mass and redshift. These merger trees are built with a mass resolution of $M_{\rm res}=6.684\times10^{10}{\rm M}_\odot$, corresponding to 40 particles in the \MDPL\ simulation, and well below the limiting mass (corresponding to 300 particles) which we will use in our analyses. Timesteps in the PCH branching algorithm are chosen to be sufficiently small that the probability of multiple branching is less than 1\%, the fraction of mass accreted in the form of unresolved halos is less than 1\%, and that the first order approximation made in equation~(2) of PCH is always valid. 

To compute branching rates in merger trees, the PCH algorithm utilizes the progenitor mass function in the limit of infinitesimal timesteps as predicted by extended Press-Schechter theory \citep{lacey_merger_1993}, multiplied by a corrective factor
\begin{equation}
 G(\sigma_1,\sigma_2,\delta_2) = G_0 \left({\sigma_1\over\sigma_2}\right)^{\gamma_1} \left({\delta_2\over\sigma_2}\right)^{\gamma_2},
\end{equation}
where $\sigma_i=\sigma(M_i)$, with $i=1$ corresponding to the progenitor halo, and $i=2$ the parent halo, while $\delta_2$ is the critical overdensity for collapse at the time of the parent halo.

To find the optimal values of the parameters $(G_0,\gamma_1,\gamma_2)$ we perform an \MCMC\ simulation in which these parameters are allowed to vary.

\subsection{Errors in N-body Halo Masses}\label{sec:SOerrors}

\begin{figure*}
 \begin{tabular}{cc}
 \includegraphics[width=85mm]{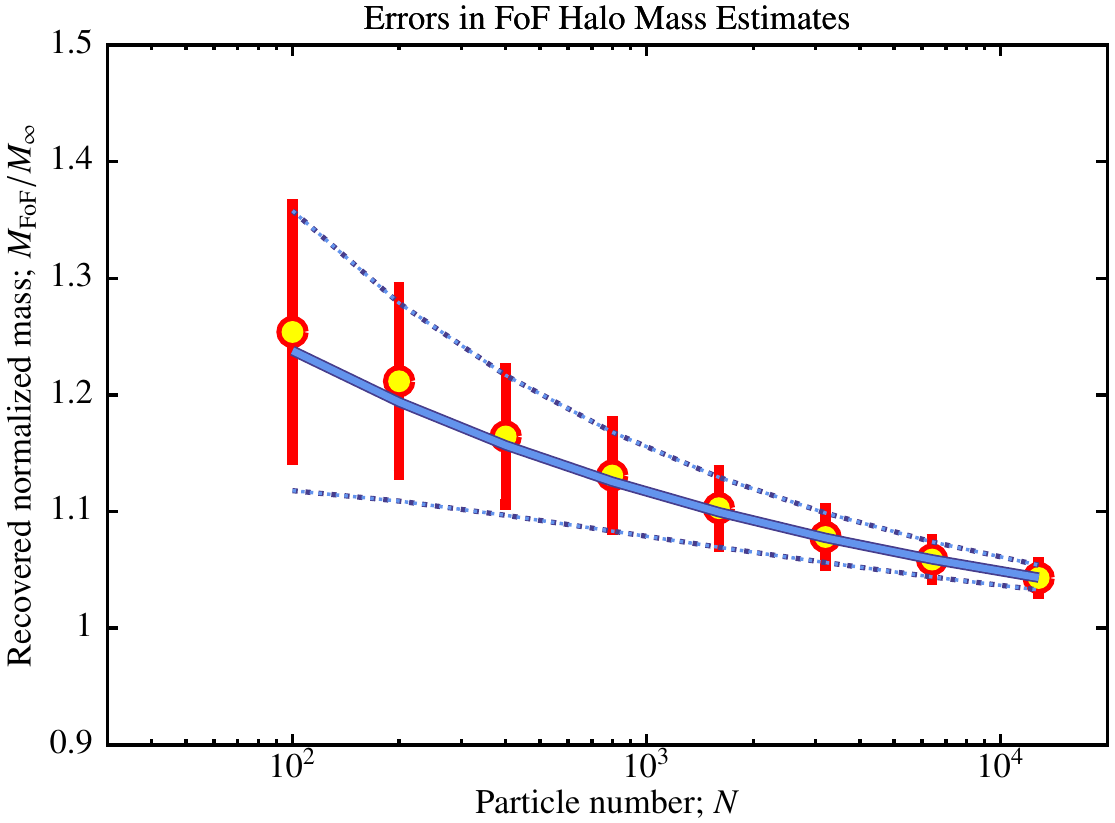} &
 \includegraphics[width=85mm]{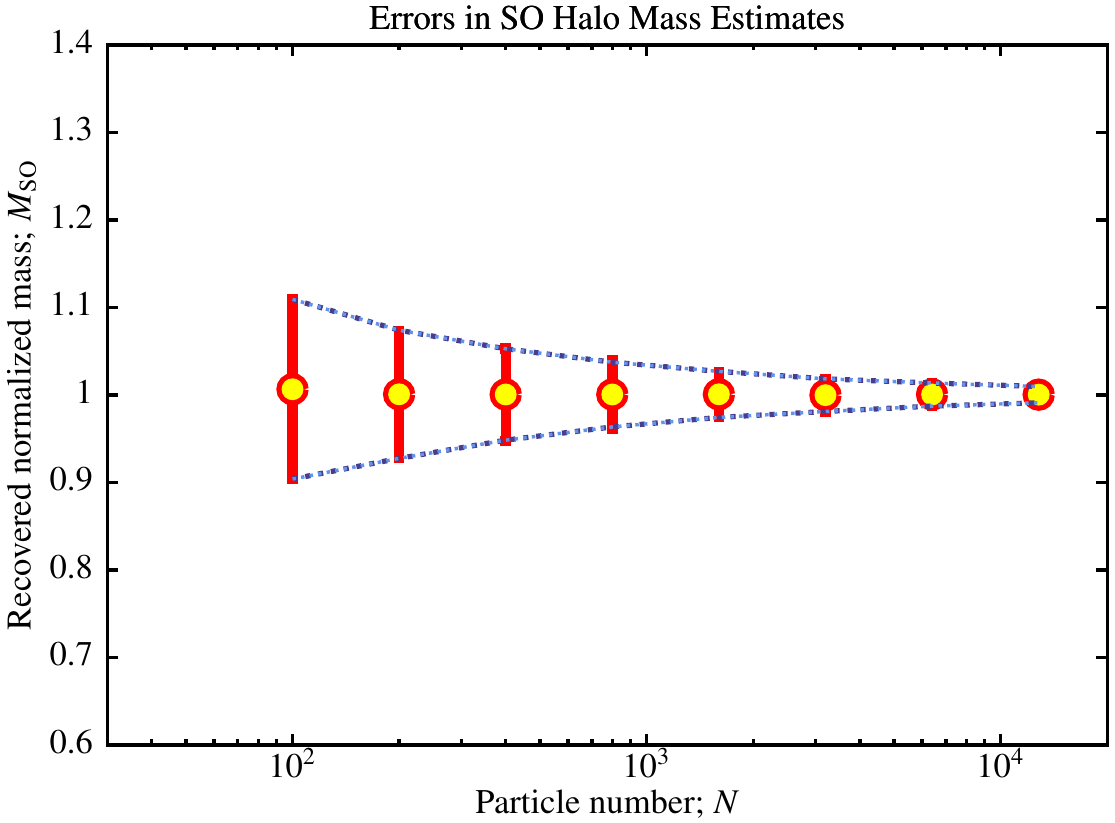}
 \end{tabular}
 \caption{The mass of N-body realizations of an idealized dark matter halo as recovered by FoF (left) and SO (right) halo finder algorithms, as a function of the number of particles (on average) in the halo. Masses are normalized to the mass that would be obtained in the limit of an infinite number of particles. Points indicate the mean mass found from a large number of halo realizations, with errorbars indicating the standard deviation in mass over this sample. The blue solid line (shown only for the FoF case) indicates the expectation for the mass based on a percolation theory analysis \protect\citep{more_overdensity_2011}. Dotted blue lines indicate models for the standard deviation in halo mass, as described in the text.}
 \label{fig:haloFinderErrors}
\end{figure*}

Dark matter halos in N-body simulations consist of a number of particles, which represent a random sampling of the underlying dark matter distribution function in the halo. This finite particle representation will inevitably lead to errors in measured properties of the halos. In this work we are concerned with halo masses. To estimate the uncertainties in halo masses arising from finite numbers of particles we perform a simple experiment. Namely we simulate a large number of spherical, isothermal (i.e. with density profiles $\rho\propto r^{-2}$) dark matter halos represented with different numbers of particles. We then apply friends-of-friends (\citealt{davis_evolution_1985}; FoF) and spherical overdensity (\citealt{lacey_merger_1994}; SO) halo finder algorithms to determine the mass of each realization. From these results we find both the mean and scatter in the recovered halo mass as a function of the number of particles used. Figure~\ref{fig:haloFinderErrors} shows the results for both halo finder algorithms. As is well known (and which can be predicted from percolation theory and is shown by the solid blue line in the left panel of Figure~\ref{fig:haloFinderErrors}; \citealt{more_overdensity_2011}) the FoF algorithm returns halo masses biased high at low numbers of particles. The SO algorithm does not (at least in this simple case, by construction). We find both algorithms show significant random errors in recovered halo masses. In the case of the FoF algorithm, we find that the standard deviation of these random errors can be described by the simple relation $\sigma(N) \approx 1.2 N^{-1/2}$, where $N$ is the mean number of particles in the halo.

In the case of the SO algorithm, the standard deviation of the random errors on halo mass can be derived analytically. Consider a sphere of fixed radius, corresponding to the radius of a halo (as defined by the SO algorithm density threshold in the limit of an infinitely well-resolved halo), centred on the location of that halo. We expect the number of particles within this sphere to be a random variable, drawn from a Poisson distribution with mean $\langle N \rangle = M/m_{\rm p}$, where $M$ is the true mass of the halo, and $m_{\rm p}$ is the particle mass of the simulation. If the number of particles inside the sphere fluctuates high the density is increased and the SO algorithm must grow the radius larger to reach its specified density threshold. The mass within the SO sphere is therefore raised even further. The opposite happens if $N$ fluctuates low. The variance in halo mass is therefore expected to be larger than that of a Poisson distribution with mean $\langle N \rangle$. Specifically, if we consider the change in the radius of the halo necessary to return to the same mean enclosed density when the mass inside the original sphere fluctuates by $\delta M$ we find that the additional mass gained by this fluctuation in radius is (to first order in $\delta M$):
\begin{equation}
 \delta^\prime M = \delta M \left[ {\rho_{\rm SO} \over \rho_{\rm s}}-1\right]^{-1},
\end{equation}
where $\rho_{\rm SO}$ is the density threshold in the SO algorithm, and $\rho_{\rm s}$ is the density at the surface of the unperturbed sphere.


If $\delta M$ has root-variance $\sigma$ we can find the root-variance of $\delta M^\prime$, $\sigma^\prime$, from the above. These two contributions to the total root variance in $M$ can be added linearly (not in quadrature as is usual with root-variances because, in this model, the two contributions are perfectly correlated). The total fractional root-variance in $M$ is then
\begin{equation}
 \sigma_{\rm total} = \sigma \left\{ 1 + \left[ {\rho_{\rm SO} \over \rho_{\rm s}}-1\right]^{-1} \right\}.
 \label{eq:SOerror}
\end{equation}
For an isothermal halo ($\rho_{\rm SO}/\rho_{\rm s}=3$), and for our idealized simulation the variance is that of a binomial distribution with $p=1/2$, namely $\sigma = 1/\sqrt{2N}$. The resulting root-variance, $\sigma_{\rm total}$ is shown by the blue dotted line in Figure~\ref{fig:haloFinderErrors} and agrees extremely well with the root-variance measured directly from our idealized simulations.

\begin{figure}
 \includegraphics[width=85mm]{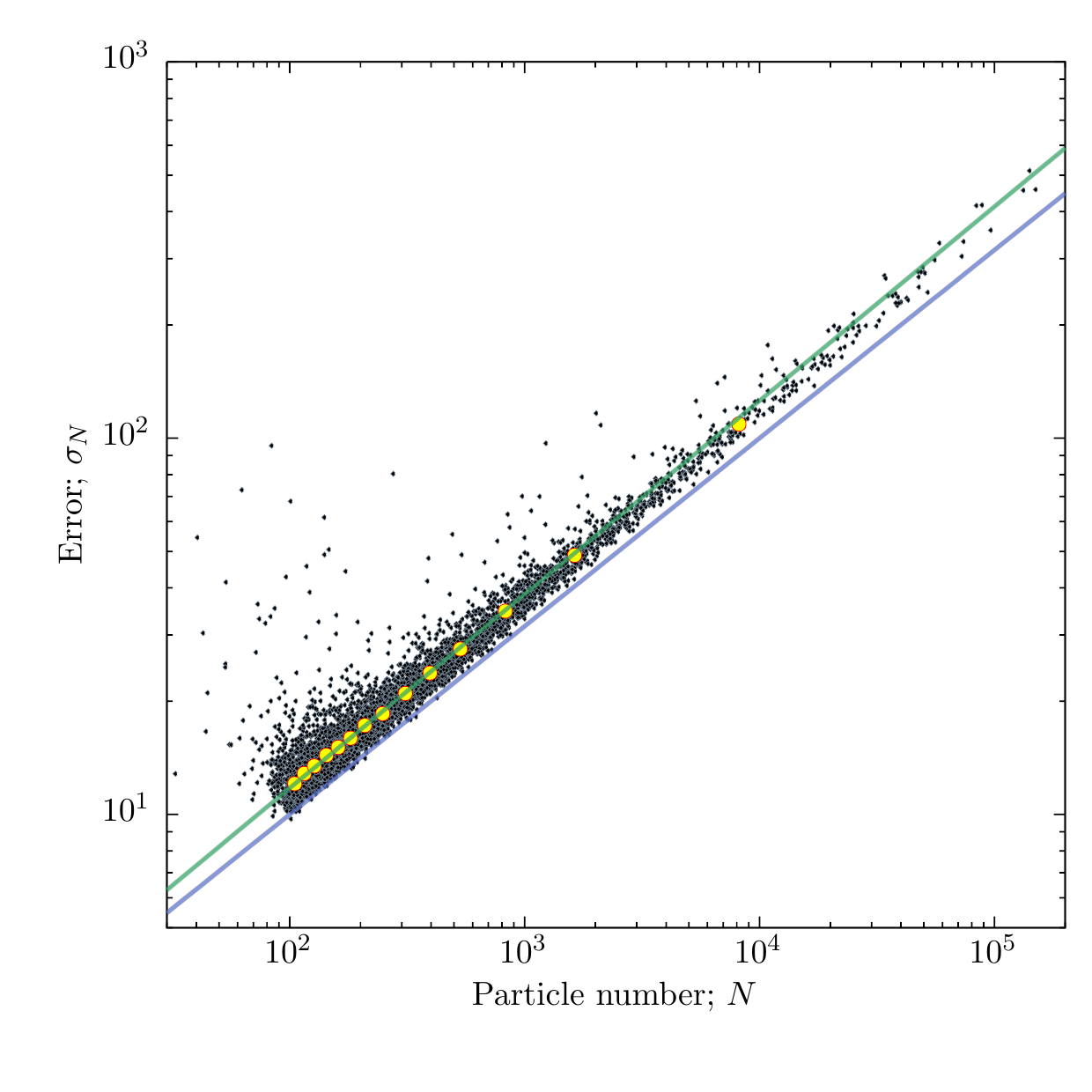}
 \caption{Errors in SO halo masses computed by resampling of N-body simulation halos. Small black points are estimates of the error in the individual SO halo mass of each simulated halo computed by resampling (with replacement) of the halo's particles. The particle number on the $x$-axis corresponds to the mean number of particles in the halo over all resamplings. Yellow circles with error bars are binned estimates of the mean error on halo mass (with 400 halos per bin), with error bars showing the standard error on the mean. The blue line shows the Poisson error estimate (i.e. $\sqrt{N}$), while the green line shows the error predicted by equation~(\protect\ref{eq:SOerror}). This clearly demonstrates both that the error in SO halo masses exceeds a simple Poisson expectation, and that our error model performs well in predicting the measured errors.}
 \label{fig:haloFinderSimulationErrors}
\end{figure}

To test how well this error model performs on actual N-body simulations we extract a sample of several thousand dark matter halos from the Millennium simulation \citep{springel_simulations_2005}, identified using the friends-of-friends algorithm, and spanning a wide range of masses (corresponding to particle numbers of around $10^2$ to $10^5$). For each halo, we extract all particles within a cubic region of length 6~Mpc/$h$ (sufficient to more than capture the entirety of the halo plus a significant region around it) centred at the halo centre of mass. We then resample the $N$ particles in this region, with replacement, to produce a new sample of $N$ particles and compute the mass of the resampled halo using the SO algorithm---a similar bootstrapping approach has recently been used by \cite{poveda-ruiz_quantifying_2016} to study the errors and biases in N-body determinations of halo concentrations. This process is repeated 1000 times for each halo, and the mean number of particles, and variance in that number is computed from these realizations. Figure~\ref{fig:haloFinderSimulationErrors} shows the resulting error on particle number for each halo (small black points), together with the mean error as a function of the number of particles contained in the halo (adaptively binned such that each point corresponds to 400 halos; yellow points). Note that the particle number plotted is the mean number found by the SO algorithm---in some instances this can be lower than the corresponding friends-of-friends halo mass and so some points extend below $N=100$. The blue line indicates Poisson (i.e. $\sqrt{N}$) errors, while the green line indicates errors predicted by equation~(\ref{eq:SOerror}) assuming \cite{navarro_universal_1997} profiles with a \cite{gao_redshift_2008} concentration-mass relation to compute the $\rho_{\rm SO} / \rho_{\rm s}$ term, and assuming $\sigma=1/\sqrt{N}$ as appropriate if the number of particles within the true halo radius (i.e. that which would be obtained in the limit of an infinite number of particles) is Poisson distributed. 

\begin{figure}
 \includegraphics[width=85mm]{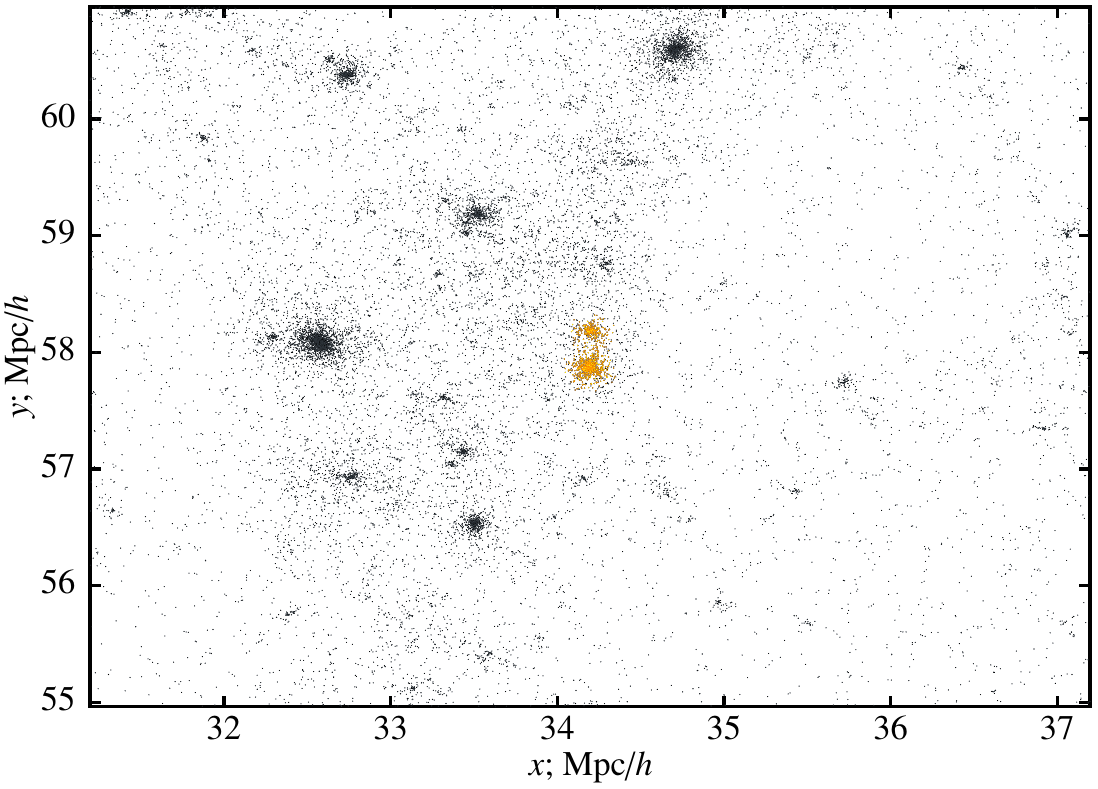}
 \caption{Particle content in (orange points) and around (black points) a FoF identified halo for which an exceptionally high mass uncertainty was determined by bootstrap resampling of its particles and applying the SO algorithm. The halo is clearly a linked group of particles in two halos, and is therefore prone to large mass fluctuations depending on precisely where the group centre is determined to lie, and how particles are sampled from the underlying continuum of dark matter.}
 \label{fig:highErrorGroup}
\end{figure}

It is clear that the Poisson model (blue line) underpredicts the measured errors in halo mass, while the model specified by equation~(\ref{eq:SOerror}) performs much better at predicting the measured errors. We note that some of the individual halo errors (shown as small black points) deviate substantially from the predicted line---this deviation is biased to higher errors. We find that in these cases the halo being resampled has significant substructure nearby, or in some cases is a linked pair of halos (from the FoF algorithm) as shown in Fig.~\ref{fig:highErrorGroup}, leading to much larger variations in the mass identified by the SO algorithm as it was resampled than predicted by our model (which assumes a spherical halo with a smooth density profile). Despite these outliers it is clear that our model performs well in predicting the error in the vast majority of cases. It is also apparent that at the highest masses (i.e. highest particle numbers) our model prediction slightly overestimates the error in halo mass. This may be because the concentration--mass relation we employ is inaccurate in this regime, or because the assumptions of our model (e.g. spherical symmetry) begin to break down. In any case, the fractional error is small in this regime, so the slight overestimate in halo mass determination error will not significantly affect our results.

If we use equation~(\ref{eq:SOerror}) to estimate the uncertainty in halos consisting of just $N=20$ particle, we predict $\sigma_N\approx 5$---a fractional error of $25\%$. Many halo catalogues (and merger trees) are constructed from halos containing such low numbers of particles---it is important to keep in mind the very large uncertainties on the resulting halo masses and how these might propagate into any predictions.

Finally, we note that our resampling procedure would be expected to give a correct estimate of the errors in halo masses if particles in N-body simulations were merely tracers of the underlying density field. In fact, while the particles play that role, they are also carriers of the density field, and so fluctuations in particle number in a given region can influence the dynamics and evolution of that region. As such, our error estimates could be biased. Furthermore, it is possible that the error in halo mass, $\sigma$, may depend on the manner in which the initial conditions of the N-body simulation were created. For example, if glass initial conditions were used the variance in the number of particles in any volume of the initial conditions will be less than that expected from a Poisson distribution. How this propagates through the to variance in the mass of halos at later times in the simulation is unclear. To address these issues ideally, one would repeat the same N-body simulation many times, with the same initial density field, but sampled by different random sets of particles, then identify the same halos in each simulation and compute the variance in their masses directly. No such suite of simulations exists, and so this experiment can not currently be performed, but could be straightforwardly be carried out in the future. Whatever the precise form of $\sigma$, it will clearly be non-zero, and our model for $\sigma_{\rm total}$ (which does not depend on the details of the distribution of the number of particles in the halo region) should still apply.

\section{Results}\label{sec:results}

Utilizing the measured backsplashless mass functions and conditional mass functions from the \MDPL\ simulation, and the SO algorithm halo mass error model of \S\ref{sec:SOerrors}, we constrain the parameters of the \cite{sheth_excursion_2002} mass function and the PCH merger tree algorithm.

\subsection{Halo Mass Functions}

We perform an \MCMC\ simulation to estimate the posterior distribution over the parameters of the \cite{sheth_excursion_2002} halo mass function when constrained to match that measured from the \MDPL\ simulation. Specifically, we perform a differential evolution \MCMC\ simulation \citep{terr_braak_markov_2006} using 16 parallel chains. At each step of the simulation a proposed state, $S_i^\prime$, for each chain, $i$, is constructed by selecting at random (without replacement) two other chains, $m$ and $n$, and finding
\begin{equation}
 S_i^\prime = S_i + \gamma (S_m - S_n) + \epsilon,
\end{equation}
where $\gamma$ is a parameter chosen to keep the acceptance rate of proposed states sufficiently high, and $\epsilon$ is a random vector each component of which is drawn from a Cauchy distribution with median zero and width parameter set equal to $10^{-3}$ of the current range of parameter values spanned by the ensemble of chains to ensure that the chains are positively recurrent. For a multivariate normal likelihood function in $N$ dimensions the optimal value of $\gamma$ is $\gamma_0=2.38/\sqrt{N}$ \citep{terr_braak_markov_2006}. We use this as our initial value of $\gamma$, but adjust $\gamma$ adaptively as the simulation progresses to maintain a reasonable acceptance rate. The proposed state is accepted with probability $P$ where
\begin{equation}
 P = \left\{ \begin{array}{ll} 1 & \hbox{ if } \mathcal{L}(S_i^\prime) > \mathcal{L}(S_i), \\ \mathcal{L}(S_i^\prime)/\mathcal{L}(S_i) & \hbox{ otherwise,} \end{array} \right.
\end{equation}
and where $\mathcal{L}$ is the likelihood function.

The simulation is allowed to progress until the chains have converged on the posterior distribution as judged by the Gelman-Rubin statistic, $\hat{R}$ \citep{gelman_a._inference_1992}, after outlier chains (identified using the Grubb's outlier test \citep{grubbs_procedures_1969,stefansky_rejecting_1972} with significance level $\alpha=0.05$) have been discarded. Specifically, we declare convergence when $\hat{R}=1.2$.

The Gelman-Rubin convergence measure relies on the chains be initialized in an over dispersed state. The state of each chain is therefore initialized by constructing 16-point unit Latin hypercubes. We generate 100 such cubes and find the cube which maximizes the minimum ($\ell^2$-norm) distance between any two points in the hypercube. Each point in this hypercube realization is used as the initial state for a chain by associating $C_i=L_i$ where $L_i$ is the $i^{\rm th}$ coordinate of the point in the hypercube, and $C_i$ is the cumulative probability distribution of the prior on parameter $i$. The parameter values are then simply found by inverting their cumulative distributions. We choose broad, uninformative priors for the parameters which span the range of values found by previous studies, specifically, $a=[0.05,1.50]$, $p=[-1,+1]$, $A=[0.05,1.00]$.

Our likelihood function is defined as
\mathchardef\mhyphen="2D
\begin{equation}
 \log \mathcal{L} = -\frac{1}{2}\sum_i \frac{\left(\phi_i^{\rm (model)}-\phi_i^{\rm (N\mhyphen body)}\right)^2}{V_i^{\rm (N\mhyphen body)}},
\end{equation}
where $\phi^{\rm (model)}$ is the mass function resulting from the \cite{sheth_excursion_2002} fitting function after convolution with the halo mass error distribution, $\phi^{\rm (N\mhyphen body)}$ is that measured in the \MDPL, $V^{\rm (N\mhyphen body)}$ is the variance in the N-body mass functions (computed under the assumption that the number of halos in each bin obeys Poisson statistics), subscript $i$ runs over all bins in the mass functions above our imposed resolution limit corresponding to $300$ particles, and which contain at least 30 halos in the \MDPL\ simulation. In computing the model expectation we average the convolved \cite{sheth_excursion_2002} mass function across the width of each bin used in estimating the \MDPL\ mass function. This ensures that any variation in the mass function across the bin is correctly accounted for.

We find that our \MCMC\ chains converge after approximately 100 steps. We allow the simulation to run for approximately 10,000 further steps, and measure a correlation length in each parameter of around 10 steps, leaving us with approximately 16,000 independent samples from the posterior distribution over the parameters.

\begin{figure*}
 \begin{tabular}{cc}
 \includegraphics[width=85mm]{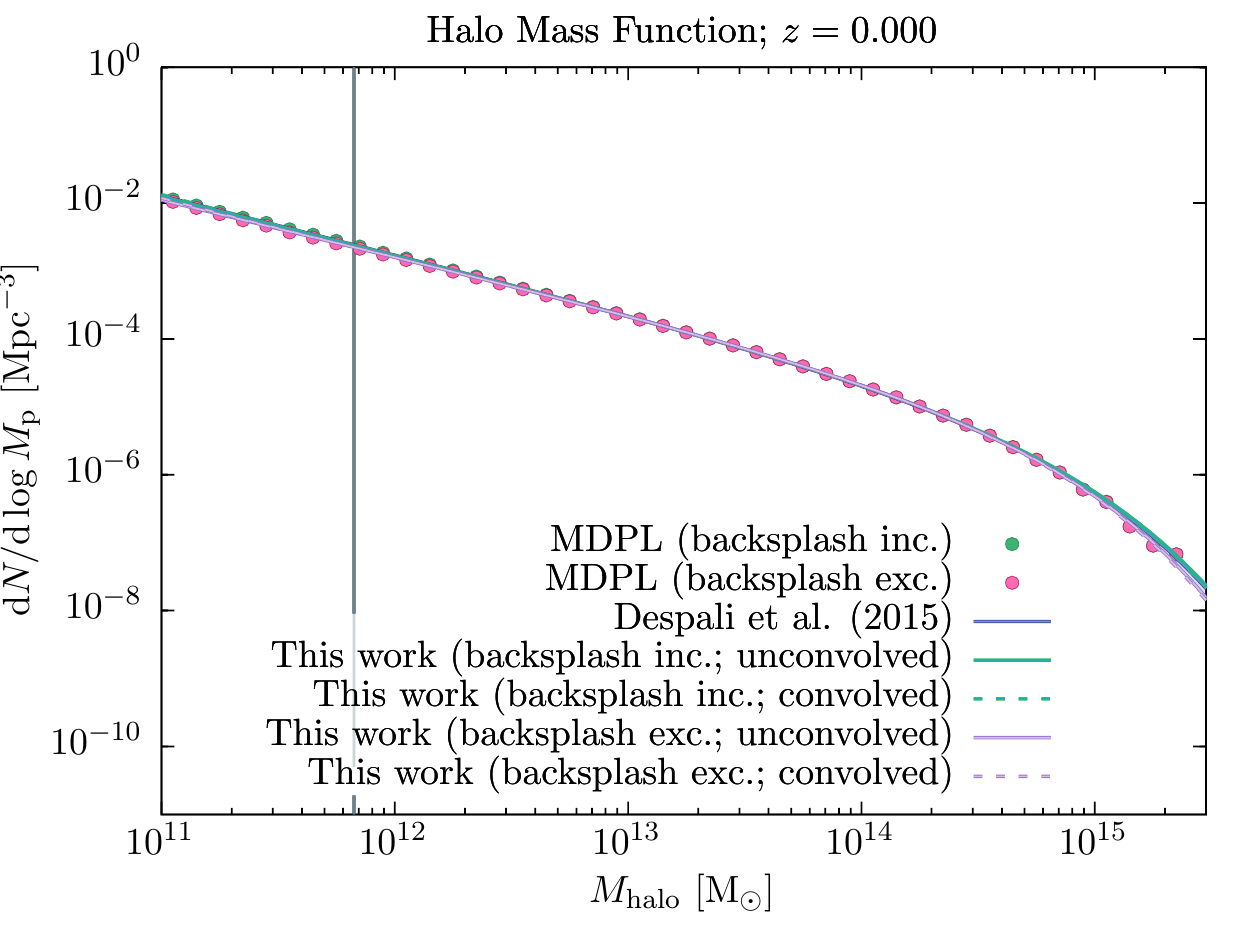} &
 \includegraphics[width=85mm]{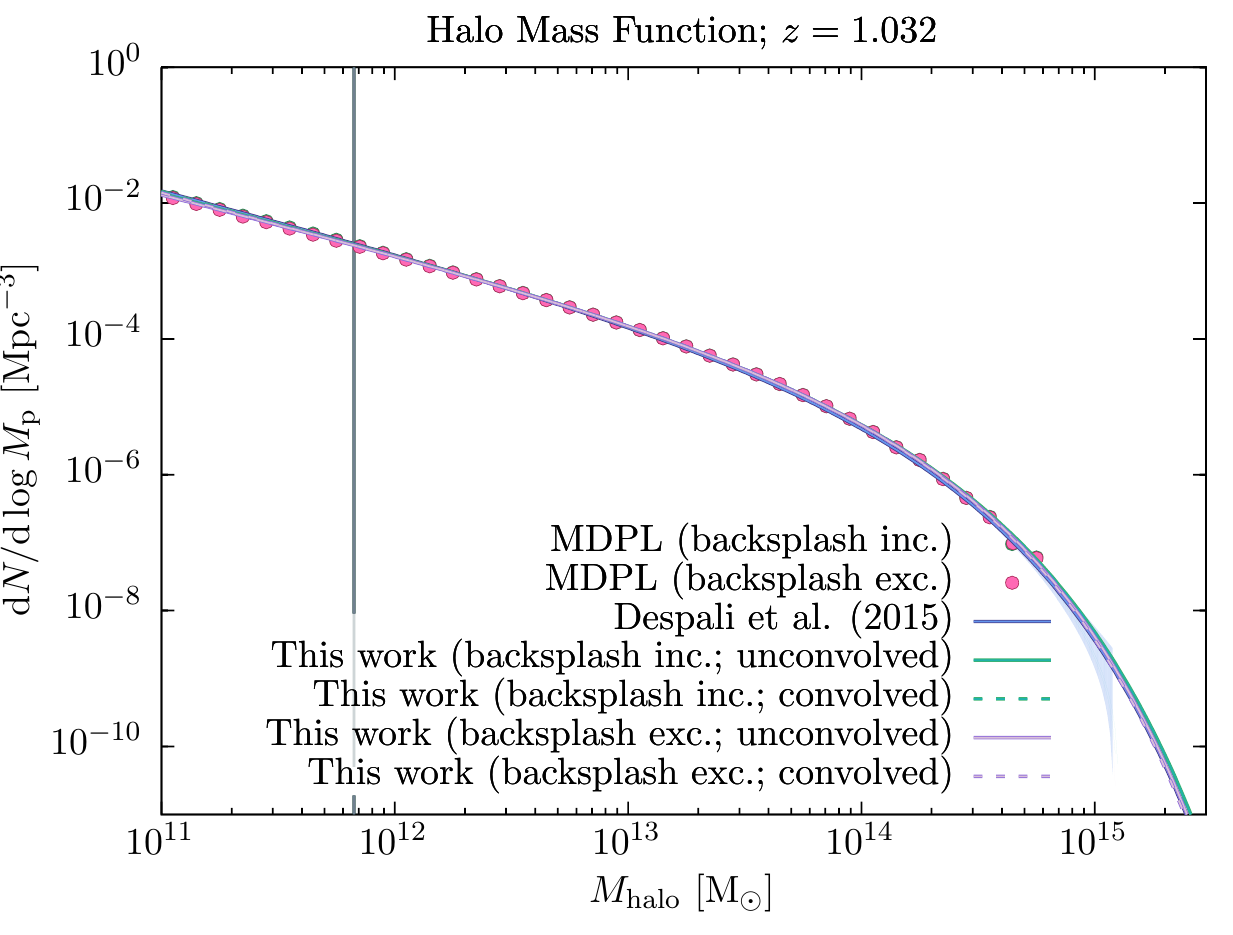} \\
 \includegraphics[width=85mm]{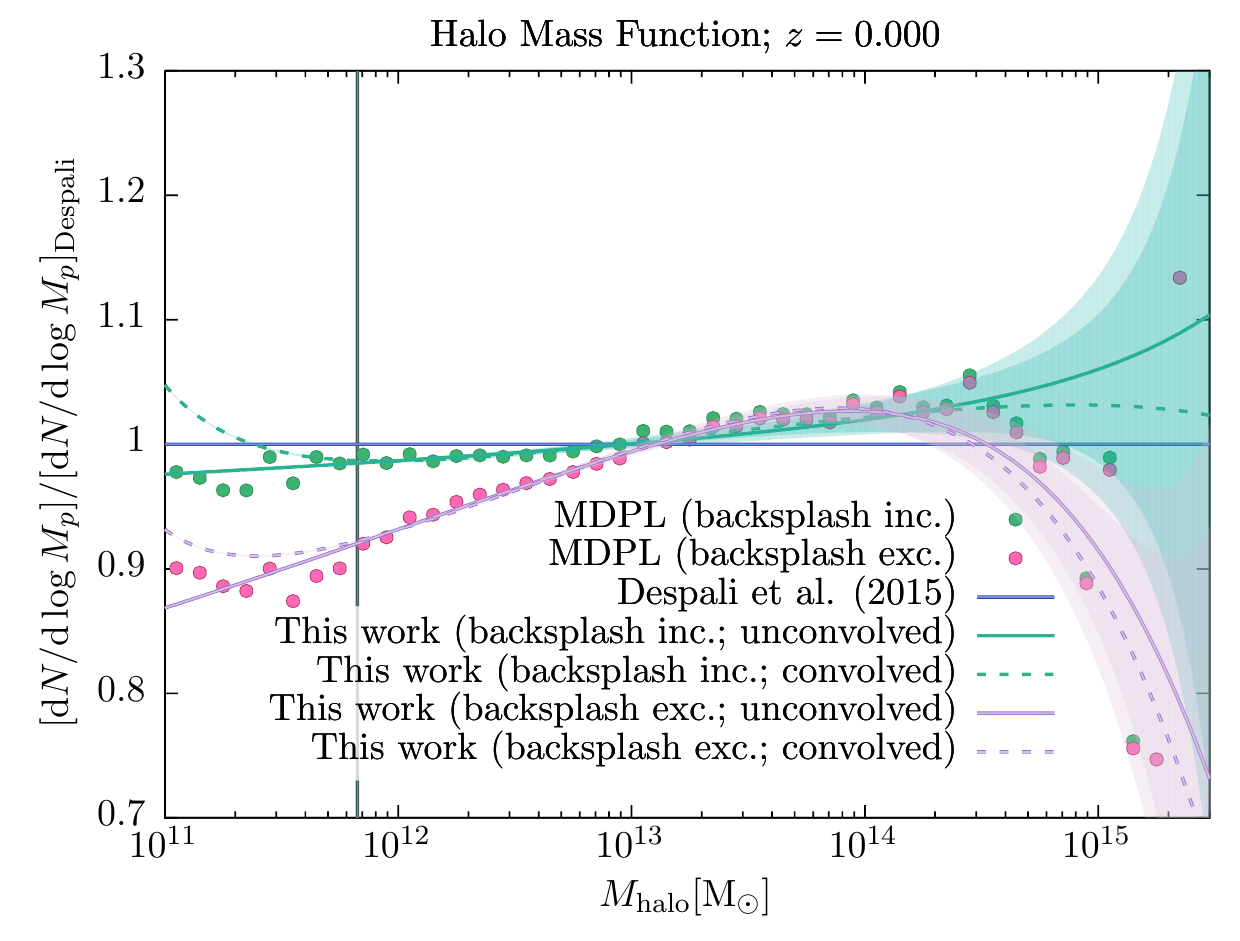} &
 \includegraphics[width=85mm]{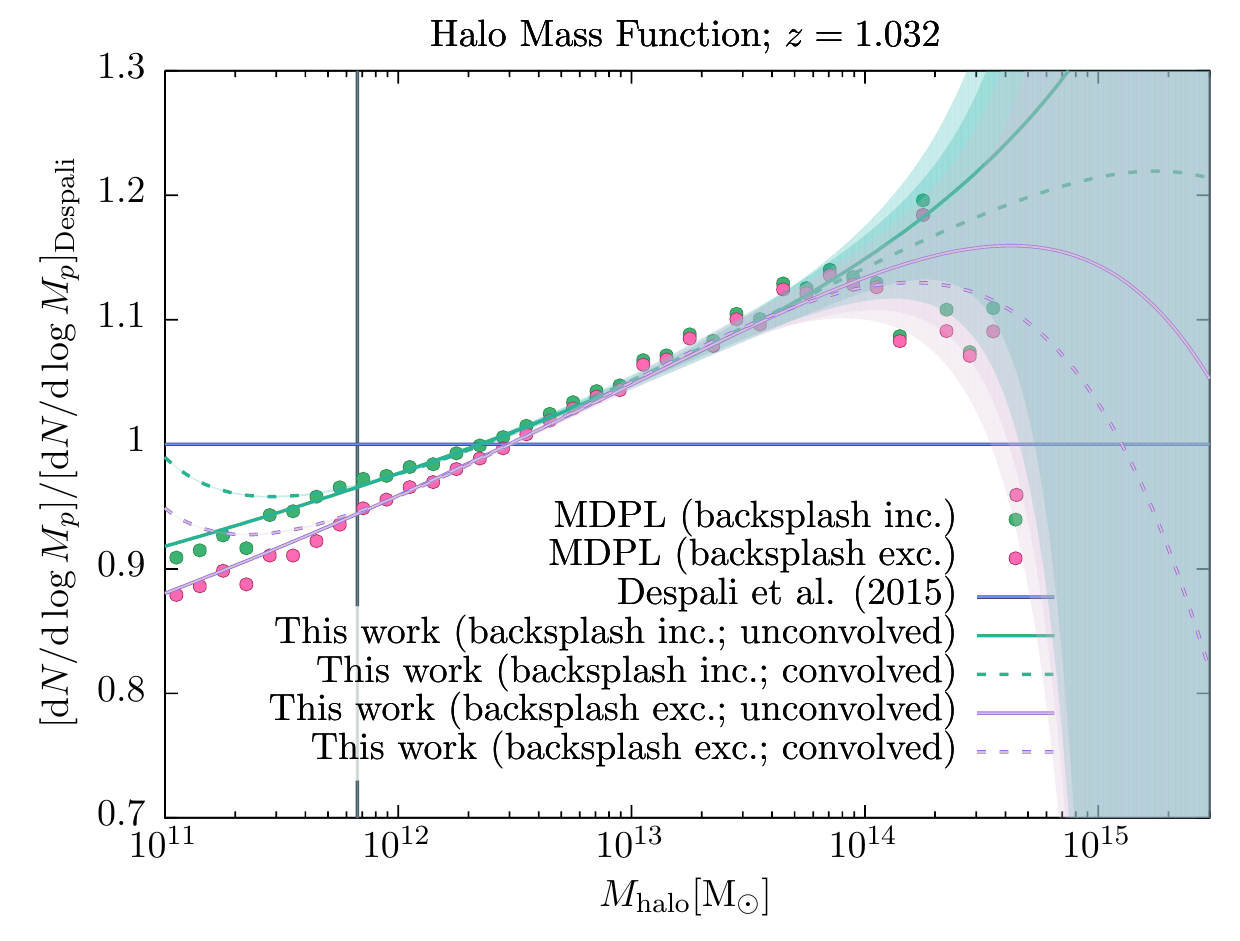} \\
 \end{tabular}
 \caption{Halo mass functions at $z=0.000$ (left column) and $z=1.032$ (right column). In the upper row we show mass functions measured from the \protect\MDPL\ simulation including backsplash halos (green points) and with backsplash halos removed (purple points). The solid blue line indicates the best fit mass function reported by \protect\cite{despali_universality_2015}, while the green and purple lines indicate the best fit mass functions to our measurements from the \protect\MDPL\ (with backsplash halos included and excluded respectively) using the fitting function of \protect\cite{sheth_excursion_2002}, convolved by the expected N-body halo mass error distribution. Transparent, shaded bands around each line indicate the expected Poisson errors in the N-body data. (Note that the fitting function from \protect\cite{despali_universality_2015} is \emph{not} convolved with the error distribution.) The lower row shows the same set of mass functions but relative to the best fit mass function of \protect\cite{despali_universality_2015} to highlight the small but significant differences between the various results. Vertical grey lines indicate the mass corresponding to 300 particles---the threshold below which we do not consider halos in this work.}
 \label{fig:massFunctions}
\end{figure*}

Figure~\ref{fig:massFunctions} shows mass functions at $z=0$ and $z\approx 1$ measured from the \MDPL\ simulation with and without the inclusion of backsplash halos (points), with the best fitting \cite{sheth_excursion_2002} mass function (after convolution by the error model of \S\ref{sec:SOerrors}) indicated by lines. For reference we also plot the mass function found by \cite{despali_universality_2015}. Table~\ref{tb:massFunctionParameters} lists the parameters of the best fit \cite{sheth_excursion_2002} mass function at $z=0$, for cases where backsplash halos are included and excluded (along with results for other redshifts, and for cases where the N-body halo mass error distribution is not accounted for). For the case with backsplash halos included the best fit parameters we find are close to, but significantly different from, those found by \cite{despali_universality_2015}. 

When backsplash halos are excluded the mass function is changed significantly at low masses. This effect is more pronounced at $z=0$ than at $z=1$. At $z=0$, the mass functions with and without backsplash halos begin to deviate below $10^{13}M_\odot$, corresponding closely with the value of $M_*$ (the characteristic halo mass defined by $\sigma(M) = \delta_{\rm c}$) at $z=0$. This is to be expected: higher mass halos are still in the stage dominated by growth via merging with smaller systems (and so are unlikely to have previously merged with any larger halo), while the evolution of lower mass halos is dominated by their merging into larger halos (e.g. \citealt{benson_self-consistent_2005}).

\begin{table*}
 \caption{Best fit values of the parameters of the \protect\cite{sheth_excursion_2002} mass function at $z=0$ through $z\approx 4$. Results are given for mass functions with backsplash halos both included and excluded, as indicate in the second column. For the $z=0$ mass functions results are given for cases in which the mass function was convolved with the expected N-body halo mass error distribution, and for cases where it was not, as indicated by the third column. For all other redshifts only the results when including the convolution with the error distribution are shown. Values quoted for each parameter are for the maximum posterior probability model, while quoted errors indicate the 16\% and 84\% percentiles of the posterior distribution of the parameter after marginalizing over the remaining two parameters. Rows corresponding to the ``most correct'' case (backsplash halos excluded, and fitting functions convolved with the expected error distribution) are highlighted with grey background.}
 \label{tb:massFunctionParameters}
  \begin{tabular}{lcclll}
  \hline
   {\bf         } & {\bf Backsplash    } & {\bf Convolved with     } &                 &                 &                \\
   {\bf Redshift} & {\bf halos excluded} & {\bf error distribution?} & {\boldmath $a$} & {\boldmath $p$} & {\boldmath $A$} \\
  \hline
  0.0000 & No & No & $0.7551^{+0.0039}_{-0.0036}$ & $0.2228^{+0.0036}_{-0.0041}$ & $0.3297^{+0.0004}_{-0.0003}$ \\
0.0000 & No & Yes & $0.7722^{+0.0038}_{-0.0040}$ & $0.1869^{+0.0041}_{-0.0040}$ & $0.3310^{+0.0003}_{-0.0003}$ \\
0.0000 & Yes & No & $0.8521^{+0.0046}_{-0.0045}$ & $0.0128^{+0.0049}_{-0.0051}$ & $0.3317^{+0.0002}_{-0.0002}$ \\
\rowcolor[gray]{0.9} 
0.0000 & Yes & Yes & $0.8745^{+0.0047}_{-0.0048}$ & $-0.0306^{+0.0053}_{-0.0054}$ & $0.3318^{+0.0002}_{-0.0002}$ \\
1.0320 & No & Yes & $0.7722^{+0.0035}_{-0.0035}$ & $-0.0127^{+0.0086}_{-0.0087}$ & $0.3114^{+0.0002}_{-0.0002}$ \\
\rowcolor[gray]{0.9} 
1.0320 & Yes & Yes & $0.8025^{+0.0039}_{-0.0040}$ & $-0.1256^{+0.0094}_{-0.0089}$ & $0.3031^{+0.0003}_{-0.0003}$ \\
2.0280 & No & Yes & $0.8003^{+0.0049}_{-0.0045}$ & $-0.2237^{+0.0162}_{-0.0168}$ & $0.2728^{+0.0013}_{-0.0014}$ \\
\rowcolor[gray]{0.9} 
2.0280 & Yes & Yes & $0.8069^{+0.0047}_{-0.0049}$ & $-0.2591^{+0.0168}_{-0.0159}$ & $0.2666^{+0.0014}_{-0.0013}$ \\
4.0380 & No & Yes & $0.8515^{+0.0118}_{-0.0137}$ & $-0.5107^{+0.0745}_{-0.0611}$ & $0.1990^{+0.0126}_{-0.0103}$ \\
\rowcolor[gray]{0.9} 
4.0380 & Yes & Yes & $0.8492^{+0.0106}_{-0.0146}$ & $-0.4938^{+0.0810}_{-0.0548}$ & $0.2009^{+0.0136}_{-0.0091}$ \\

  \hline
 \end{tabular}
\end{table*}

\begin{figure}
 \includegraphics[width=85mm]{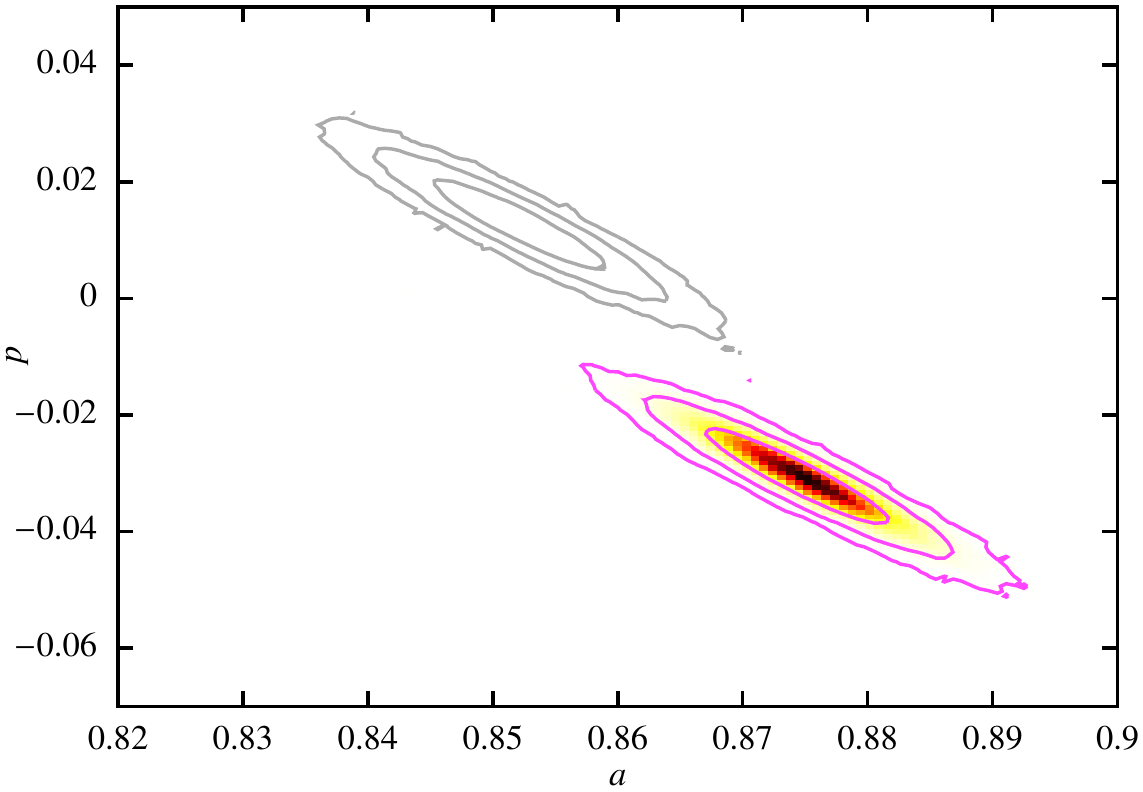}
 \caption{Constraints on mass function parameters, $a$ and $p$. The colour shading and purple contours show results for $z=0.0$ with backsplash halos excluded and with the mass function convolved with the expected distribution of N-body halo mass errors. The grey contours indicate constraints obtained ignoring errors in halo masses clearly demonstrating a significant offset.}
 \label{fig:massFunctionConstraints}
\end{figure}

The importance of convolving the fitting function with the expected N-body halo mass error distribution is illustrated in Figure~\ref{fig:massFunctionConstraints} where we show the constraints on the parameters $a$ and $p$ derived from our \MCMC\ simulation. The colour shading and purple contours indicate the posterior distribution over these parameters when fitting a convolved mass function to the N-body data, while the grey contours indicate the posterior distribution obtained if errors on N-body halo masses are ignored. Clearly, at the level of precision that can be obtained from state-of-the-art N-body simulations the treatment of errors has a very significant (much greater than $3\sigma$) effect on the resulting parameters of fitting functions.

We use \PPCs\ \citep{gilks_w._r._markov_1995,gelman_a._bayesian_2013} to test whether the model family characterized by the \PPD\ is a viable description of the N-body data. We adopt a test-statistic (or ``discrepancy'') of:
\begin{equation}
 \mathcal{T}_l = \mathcal{T}(\boldsymbol{\phi}_l) = \Delta_l\cdot\mathbfss{C}^{-1}\cdot\Delta_l^{\rm T},
 \label{eq:testStatistic}
\end{equation}
where $\Delta_l = \boldsymbol{\phi}_l - \bar{\boldsymbol{\phi}}$, $\boldsymbol{\phi}_l$ is the $l^{\rm th}$ realization of the N-body data, $\bar{\boldsymbol{\phi}}$ is the mean over these realizations, and $\mathbfss{C}$ is the covariance matrix of the realizations. To construct a realization of the N-body data we draw a set of parameters from the \PPD\ by choosing a random state from the converged portions of our \MCMC\ chains, construct the model mass function from these parameters, use this to determine the mean number of halos of each mass (after convolution with our error model) present in a volume equal to that of the \MDPL, draw a realization of the number of halos from a Poisson distribution with these means, and finally convert this back to a mass function by dividing through by the simulation volume.

We compute the same quantity for the observational data
\begin{equation}
  \mathcal{T}^\prime = \Delta^\prime \cdot\mathbfss{C}^{-1}\cdot\Delta^{\prime \rm T},
\end{equation}
where $\Delta^\prime = \boldsymbol{\phi}^\prime - \bar{\boldsymbol{\phi}}$, and $\boldsymbol{\phi}^\prime$ is the N-body data from the \MDPL. The Bayesian $p$-value is then
\begin{equation}
 \hat{p}_{\rm B} = {1\over L} \sum_{l=1}^L I_{\mathcal{T}_l \ge \mathcal{T}^\prime}.
\end{equation}
We find that $p_{\rm B}<10^{-3}$, indicating that the \cite{sheth_excursion_2002} parametric form is formally not a good description of the N-body halo mass function.

\subsection{Merger Tree Branching Rates}

To constrain branching rates of merger trees we make use of the algorithm of \cite{cole_hierarchical_2000} as modified by PCH. Specifically, we constrain the parameters $(G_0, \gamma_1, \gamma_2)$ of the PCH algorithm such that the resulting merger trees agree as closely as possible with conditional mass functions measured from the \MDPL\ simulation. For given $(G_0, \gamma_1, \gamma_2)$ we construct a large sample of merger trees spanning a range of masses, and estimate conditional mass functions from them. In constructing these conditional mass functions we convolve by the error distribution (see \S\ref{sec:SOerrors}) on both the parent and progenitor halo masses. As such, while the PCH algorithm can never intrinsically produce halos with $M_{\rm p}/M_0>1$, where $M_{\rm 0}$ is the mass of the parent halo, and $M_{\rm p}$ is the mass of the progenitor halo, it is possible to populate this region of the mass function after convolution. 

We estimate errors on the conditional mass function (both N-body and PCH) assuming each halo contributing to a bin is independent of all other halos  (i.e $\sqrt{N}$ errors). We then perform an \MCMC\ simulation to determine the posterior probability distribution over the set of parameters given the data. We adopt uninformative, uniform priors on all three parameters with ranges $G_0=[0.1,1.0]$, $\gamma_1=[-0.7,+0.7]$, and $\gamma_2=[-0.3,+0.3]$, chosen to be broad and to include the values previously found by PCH and \cite{benson_constraining_2008}. Our likelihood function is defined as
\begin{equation}
 \log \mathcal{L} = -\frac{1}{2} \Delta \cdot (\mathbfss{C}^{\rm (model)}+\mathbfss{C}^{\rm (N\mhyphen body)})^{-1} \cdot \Delta^{\rm T}
\end{equation}
where $\Delta = \phi^{\rm (model)}-\phi^{\rm (N\mhyphen body}$, $\phi^{\rm (model)}$ is the conditional mass function predicted by the PCH algorithm, $\phi^{\rm (N\mhyphen body)}$ is that measured from the \MDPL\ simulation, $\mathbfss{C}$ are the covariance matrices of the conditional mass functions (for the N-body simulation data we find $\mathbfss{C}^{\rm (N\mhyphen body)}$ by assuming that the number of halos in each bin obeys Poisson statistics, and that bins are independent, while for the model we take into account correlations between bins which arise because the mass functions are convolved with the expected N-body mass error distribution), subscript $i$ runs over all viable bins in the conditional mass functions at all viable redshifts.

By ``viable'' we mean mass and redshifts intervals where the N-body data is reliable, and the PCH model is able to provide a good description of the N-body data. Specifically, we exclude all points below the mass resolution threshold of the N-body simulation (corresponding to 300 particles). Also, while our convolved PCH conditional mass functions do populate the region $M_{\rm p}/M_0>1$ there is no reason to think that they should actually be a good description of the N-body data in this regime. That would be true only if the N-body merger trees only ever populated this region due to mass errors. In reality there may be true cases of $M_{\rm p}/M_0>1$ (e.g. after major mergers). Therefore, we exclude points in the conditional mass functions at mass ratios larger than that corresponding to the peak (if any) in the conditional mass function. Finally, we exclude conditional mass functions at $z<0.4$---based on initial explorations we find that the PCH model is unable to adequately describe the N-body data in this regime. We comment further on this below.

We find that our \MCMC\ simulation converges after 70 steps. We allow it to run for a further 140 steps, generating a total of 2,240 post-convergence states. The correlation length in our chains is approximately 40 steps, so this leaves us with only 56 independent post-convergence states. This is a rather low number---limited by the high computational cost of each model evaluation---but sufficient to at least approximately characterize the posterior probability distribution over our parameters.

\begin{figure}
 \newcommand{\triangledir}{.}
\renewcommand{\arraystretch}{0}
\setlength{\tabcolsep}{0pt}
\begin{tabular}{l@{}l@{}l@{}}
\includegraphics[scale=0.95]{\triangledir/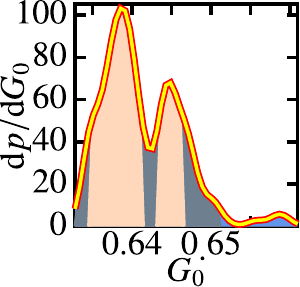}\vspace{+14.5pt}&\hspace{-2pt}\raisebox{+14.5pt}{\includegraphics[scale=0.95]{\triangledir/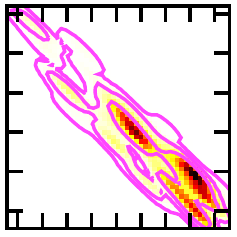}}&\hspace{-3.5pt}\raisebox{+16.0pt}{\includegraphics[scale=0.95]{\triangledir/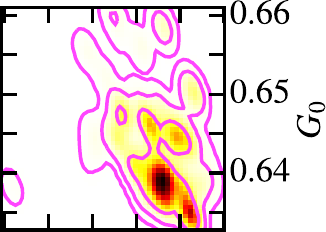}}\vspace{-33pt}\\
&\hspace{-21.5pt}\includegraphics[scale=0.95]{\triangledir/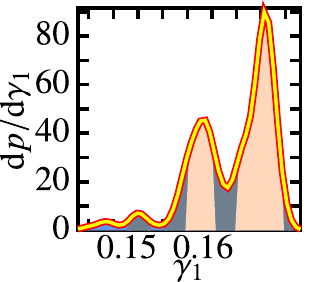}\vspace{0pt}&\hspace{-3.5pt}\raisebox{13.5pt}{\includegraphics[scale=0.95]{\triangledir/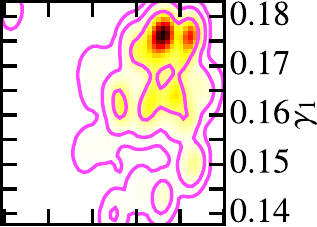}}\vspace{-14.5pt}\\
&&\hspace{-24.5pt}\includegraphics[scale=0.95]{\triangledir/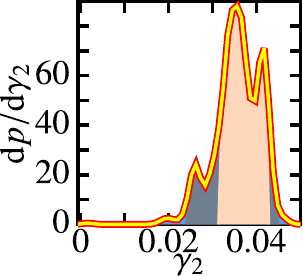}\vspace{0pt}\\
\end{tabular}

 \caption{The posterior probability distribution over parameters of the PCH algorithm. Off-diagonal panels show the posterior distribution over pairs of model parameters, while on-diagonal panels show the posterior distribution over individual model parameters. In off-diagonal panels, colours show the probability density running from white (low probability density) to dark red (high probability density). Contours are drawn to enclose 99.7\%, 95.4\%, and 68.3\% of the posterior probability when ranked by probability density (i.e. the highest posterior density intervals). In on-diagonal panels the curve indicates the probability density. Shaded regions indicate the 68.3\%, 95.4\%, and 99.7\% highest posterior density intervals.}
 \label{fig:pchTriangle}
\end{figure}

Figure~\ref{fig:pchTriangle} shows the resulting posterior distribution over the model parameters. All three parameters are constrained quite precisely, with maximum likelihood values (plus uncertainties) listed in Table~\ref{tb:pchParameters}. The joint distributions over the parameters are less well-characterized, but suggest a degree of degeneracy between $G_0$ and $\gamma_1$. Our value of $G_0$ is slightly larger than that of previous studies (PCH, \citealt{benson_constraining_2008}), while our $\gamma_1$ value is significantly lower. Perhaps most interesting though is that our preferred value for $\gamma_2$ is positive, while previous studies have found $\gamma_2<0$. As discussed by PCH, $\gamma_2>0$ results in an enhancement in the merger rate for halos with $M > M_*$. While our constrained value of $\gamma_2$ is significantly above zero, and significantly different from previous estimates, it remains intrinsically small, such that merger rates differ by only around 5\% across typical ranges of halo mass compared to the $\gamma_2=0$ case.

\begin{table*}
 \caption{Parameters of the PCH algorithm for merger tree branching rates from previous works and from this work. For this work, values quoted correspond to the maximum of the posterior likelihood, with errors corresponding to the 68.3\% highest posterior density interval.}
 \label{tb:pchParameters}
 \begin{tabular}{llll}
  \hline
  {\bf Fit}                                & {\boldmath $G_0$}             & {\boldmath $\gamma_1$}        & {\boldmath $\gamma_2$}        \\
  \hline
  PCH                                      & $+0.570$                      & $+0.380$                      & $-0.010$                      \\
  \protect\cite{benson_constraining_2008}  & $+0.605$                      & $+0.375$                      & $-0.115$                      \\
  This work                                & $+0.6353^{+0.0108}_{-0.0002}$ & $+0.1761^{+0.0023}_{-0.0153}$ & $+0.0411^{+0.0007}_{-0.0086}$ \\
  \hline
 \end{tabular}
\end{table*}

\begin{figure*}
 \begin{tabular}{cc}
  \includegraphics[width=85mm]{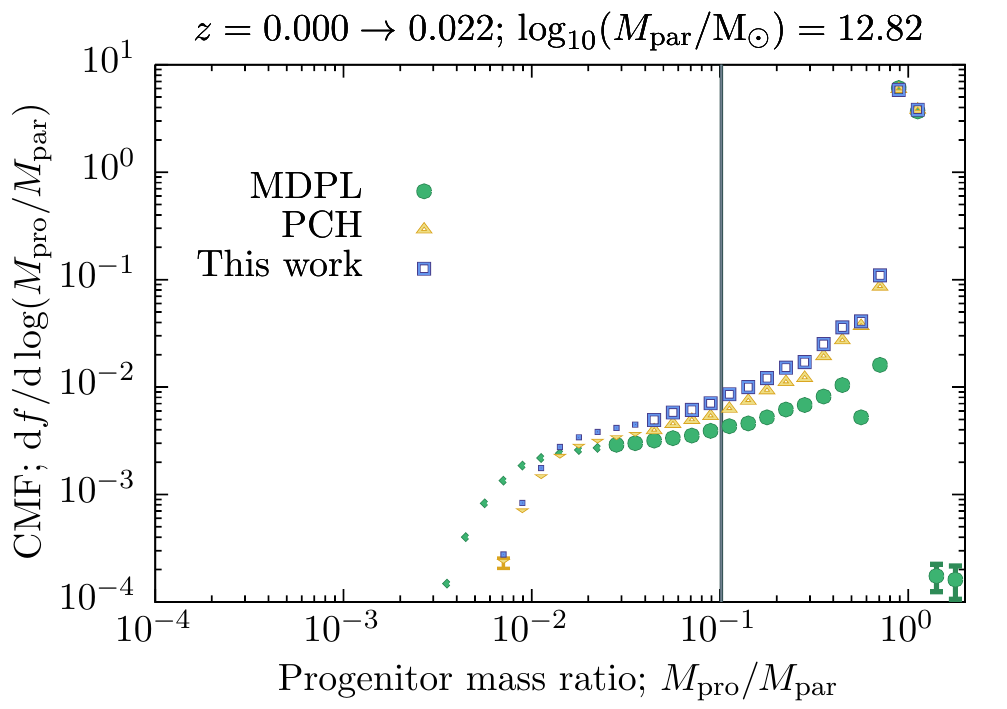} &
  \includegraphics[width=85mm]{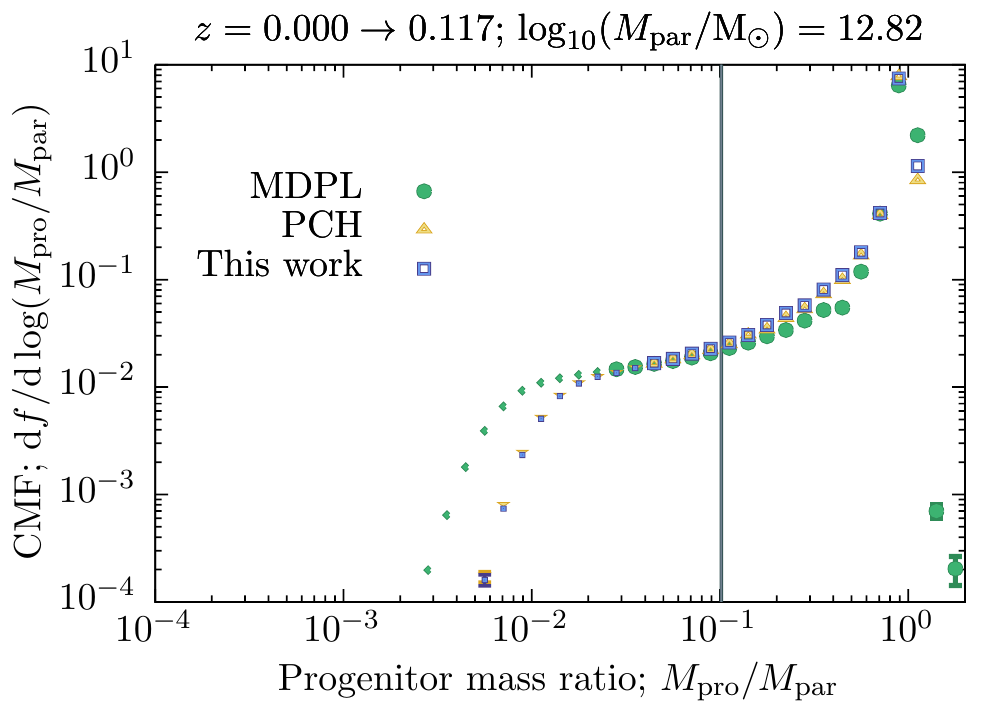} \\
  \includegraphics[width=85mm]{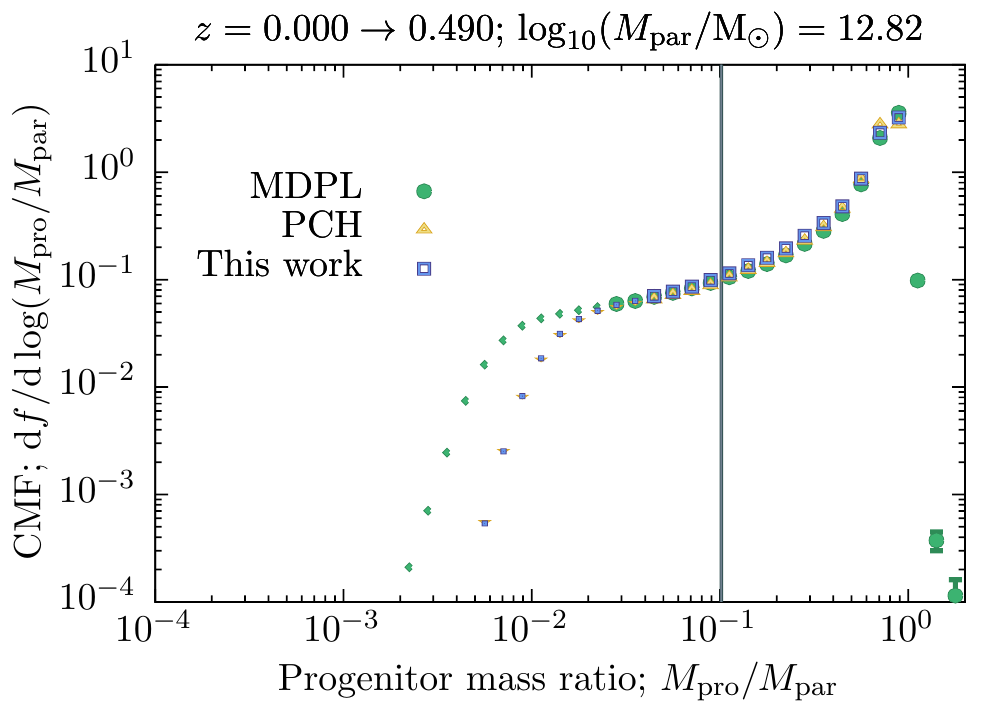} &
  \includegraphics[width=85mm]{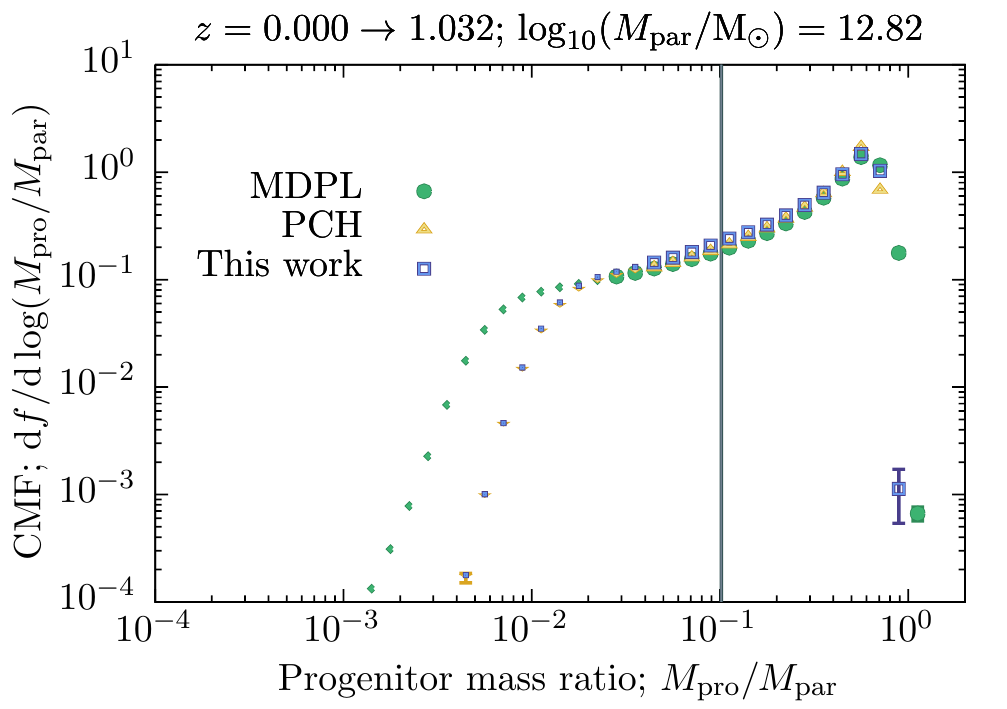} \\
  \includegraphics[width=85mm]{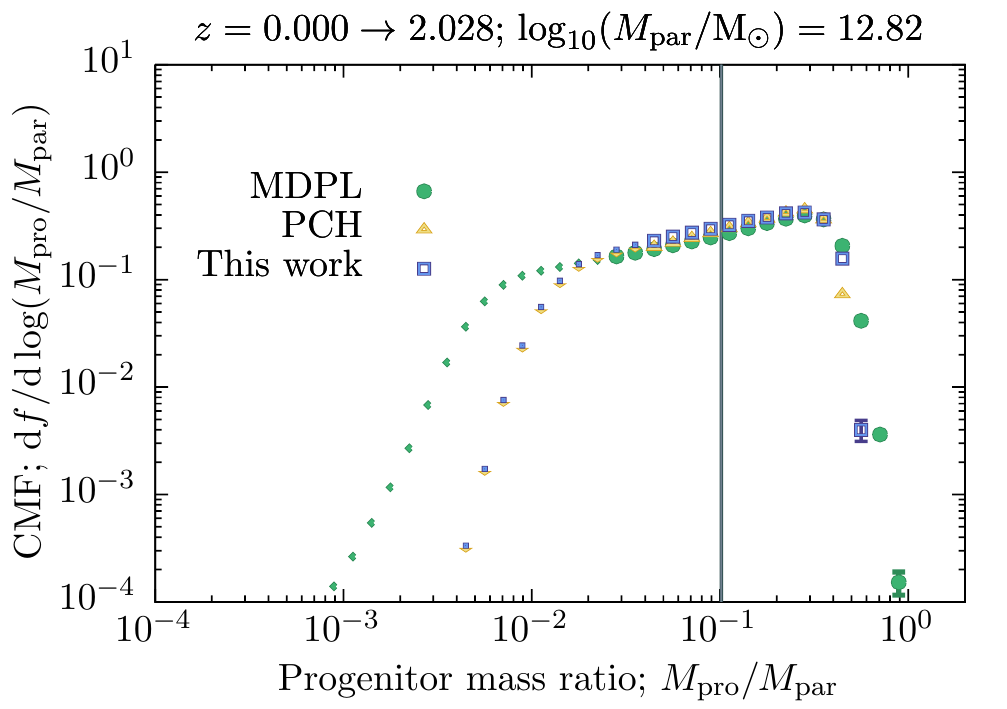} &
  \includegraphics[width=85mm]{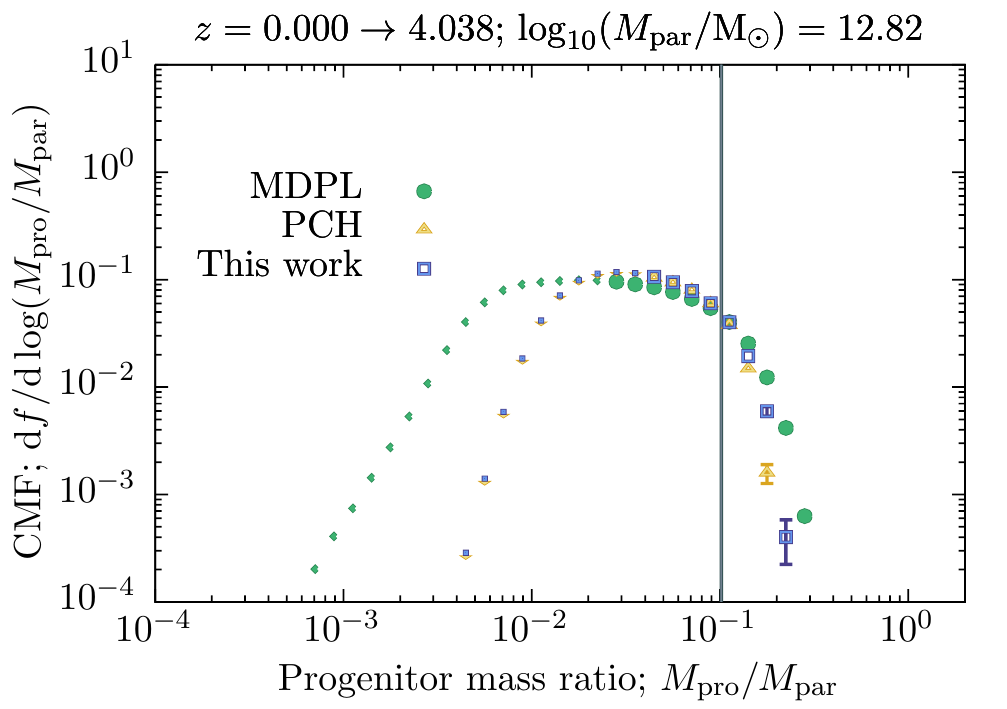} \\
 \end{tabular}
 \caption{Conditional mass functions at different redshifts (as shown above each panel) for halos of mass $\log_{10}(M/{\rm M}_\odot)=12.82$ at $z=0$. Filled green circles indicate results from the \protect\MDPL\ simulation, blue open squares indicate results from the maximum posterior likelihood model found in this work, and yellow triangles indicate results obtained using the original parameters proposed by PCH. Small symbols are used in the region where each model is affected by numerical resolution. The vertical grey line indicates the resolution limit of the simulation---points at lower mass ratios are excluded from our fitting procedure, as are points at mass ratios above the peak (if any) of the conditional mass function, and all points in conditional mass functions at $z<0.4$. Results using the PCH algorithm (including both those generated using the original PCH parameters and the parameters found in this work) are shown convolved with the expected error distributions for halo masses for both parent and progenitor halos.}
 \label{fig:cmfRedshiftSequence}
\end{figure*}

Figure~\ref{fig:cmfRedshiftSequence} shows conditional mass functions for $z=0$ parent halos in a narrow range of mass $\log_{10}(M/{\rm M}_\odot)=12.64$--$13.00$. Each panel shows the conditional mass function at a different redshift, as indicated above the panel. Filled green circles indicate results from the \MDPL\ simulation, blue open squares indicate results from the maximum posterior likelihood model found in this work, and yellow triangles indicate results obtained using the original parameters proposed by PCH. Small symbols are used in the region where each model is affected by numerical resolution. The vertical grey line indicates the mass above which points are used in our fitting procedure. Results using the PCH algorithm (including both those generated using the original PCH parameters and the parameters found in this work) are shown convolved with the expected error distributions for halo masses for both parent and progenitor halos.

Overall, the PCH algorithm performs well in matching the N-body conditional mass functions, as was demonstrated by PCH. The fitting parameters derived in this work, while a better match to the N-body data as judged by our goodness of fit metric\footnote{We find that the higher likelihood of our parameters compared to those found by PCH is largely driven by the behaviour in high mass ratio bins at higher redshifts. In this regime, both choices of parameter values underpredict the N-body results, but the values found in this work underpredict less dramatically. This, coupled with the fact that PCH constrained parameter values to match N-body results from a different (in both cosmology, halo finder algorithm, and tree construction algorithm) simulation, and did not convolve their predicted conditional mass functions with the expected N-body halo mass error distribution is the cause of the difference of our results from those of PCH.}, do not clearly perform any better than those found by PCH given the overall ability of the PCH algorithm to match the N-body results. As shown by PCH, the PCH algorithm is able to produce a reasonably good match to the shape and evolution with redshift of the conditional mass function, but fails to match some of the details. In particular, they underpredict the conditional mass function for large progenitor mass ratios, even after being convolved with the expected halo mass error distributions. This indicates that there are either actual physical processes at work which contribute to these parts of the conditional mass function, or additional numerical/definitional issues associated with halo identification not accounted for by our error model. Such behaviour has been clearly demonstrated by \cite{behroozi_major_2015} who trace a well-resolved major merger with multiple halo finders, including {\sc RockStar}. In this case (a 1:1.8 mass ratio merger) the halo mass reached around 130\% of its final value during the merging process.

The most notable failure however occurs for the conditional mass function at $z=0.022$---the smallest redshift interval that we consider (a similar, but lesser failure is visible in the conditional mass function at $z=0.117$ also). This is why we exclude these redshifts from our fitting procedure---initial exploratory studies suggested that the PCH algorithm is unable to give a good match in this regime, and so it is not meaningful to attempt to fit the model to this data. Here, the PCH algorithm tends to overproduce the number of lower mass progenitors. This redshift interval corresponds to a time interval of less than 300~Myr, less than the dynamical times of dark matter halos at $z\approx 0$ and so less than the typical timescale for mergers of these halos. This failure of the PCH algorithm may therefore reflect a breakdown in one assumption of that algorithm, namely that halo merging is an instantaneous process. 

\begin{figure*}
 \begin{tabular}{cc}
  \includegraphics[width=85mm]{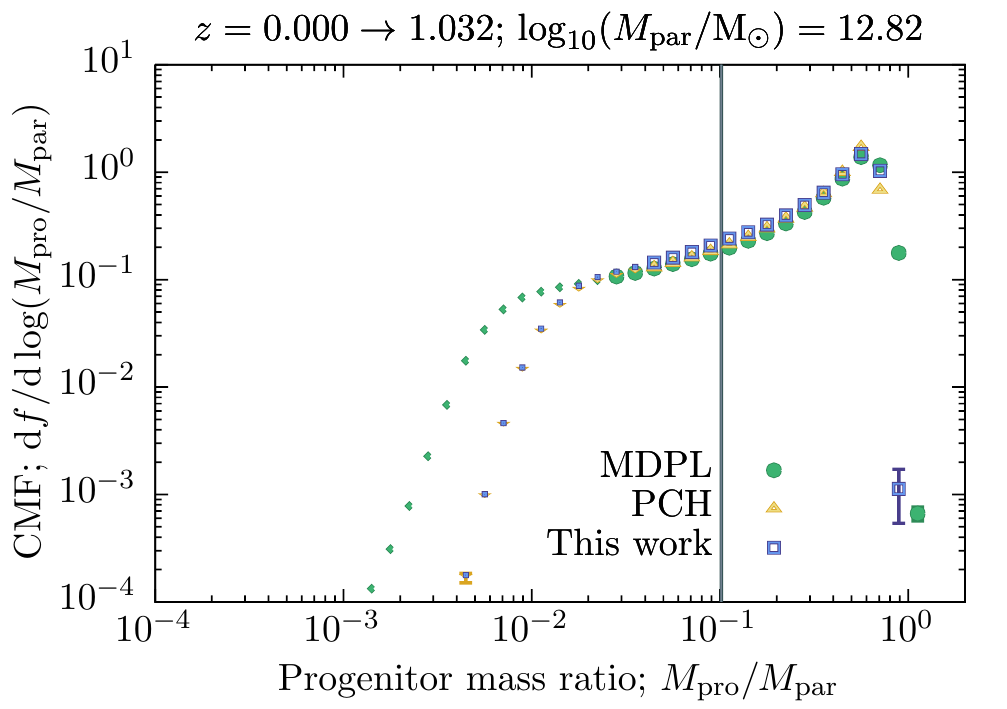} &
  \includegraphics[width=85mm]{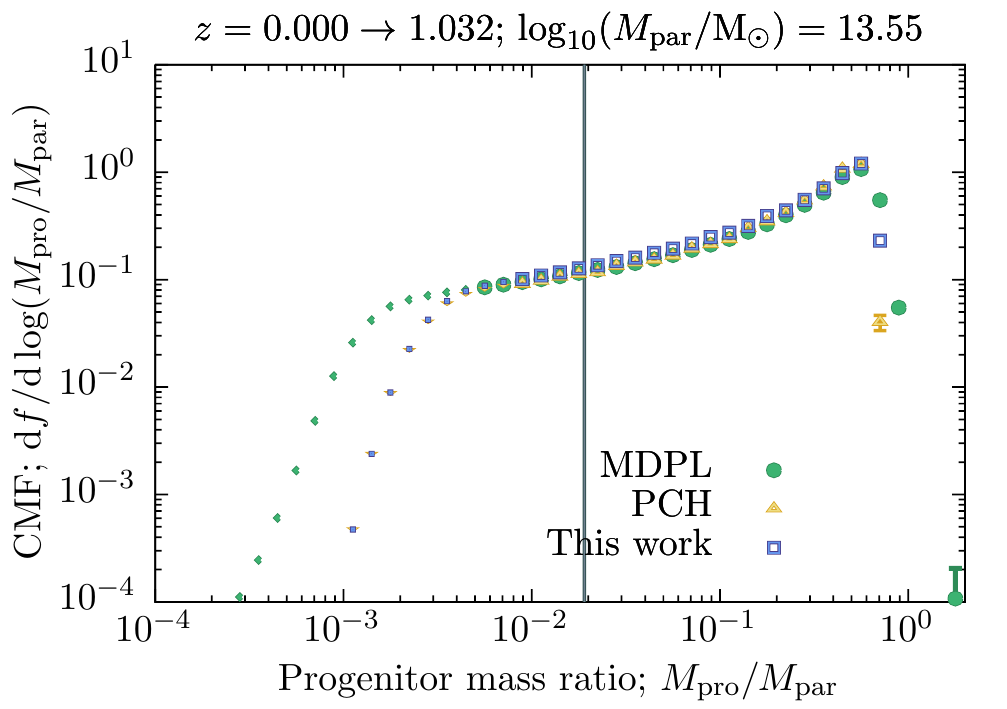} \\
  \includegraphics[width=85mm]{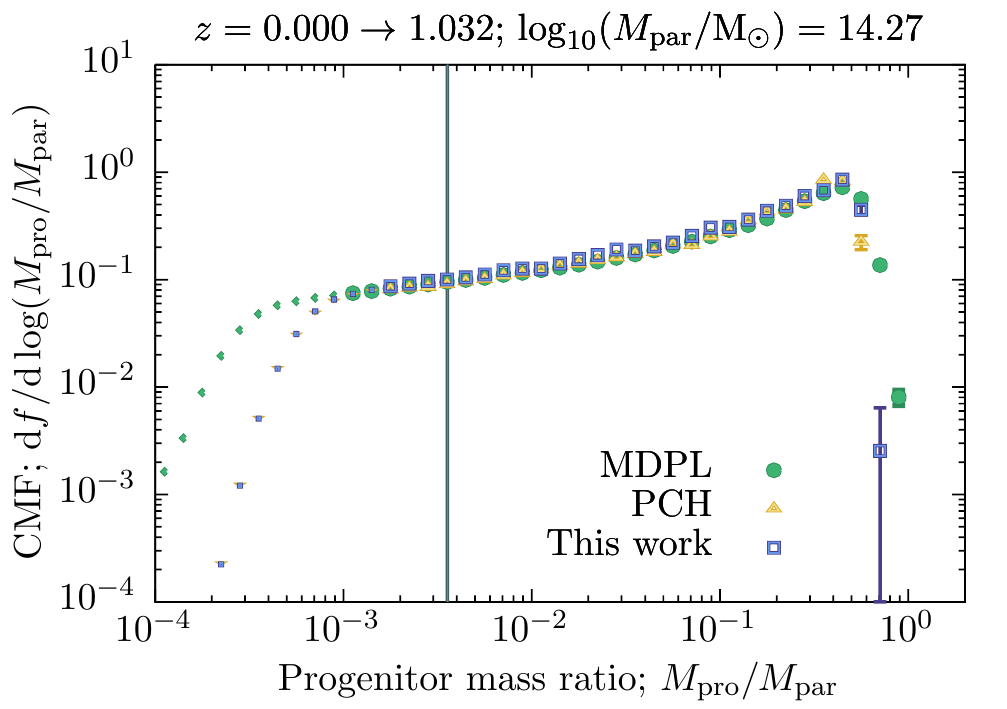} &
  \includegraphics[width=85mm]{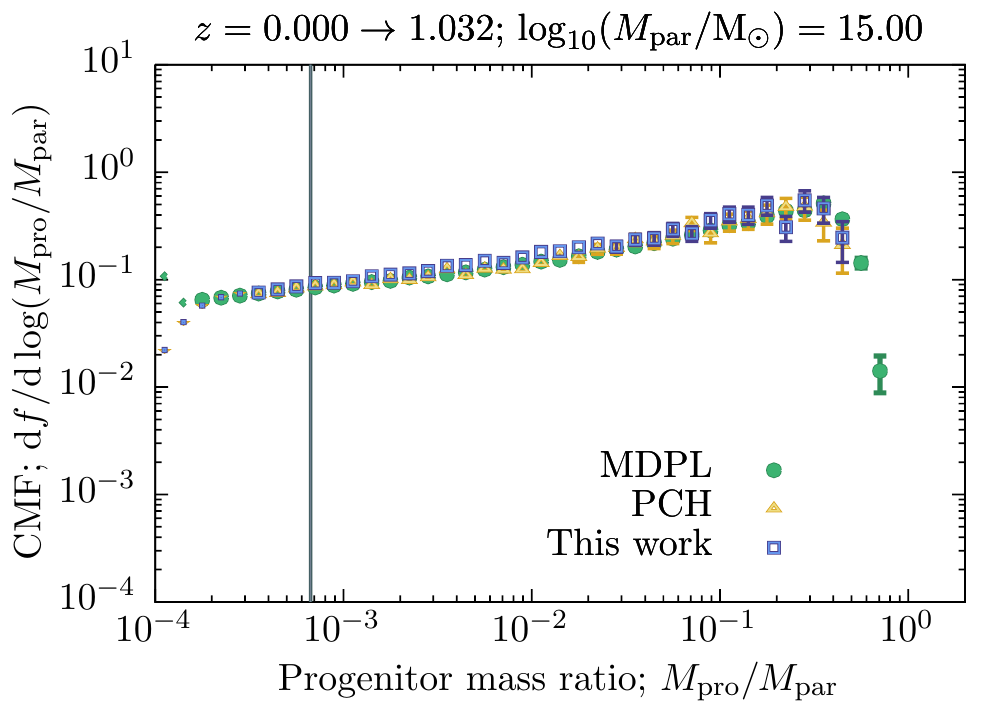} \\
 \end{tabular}
 \caption{Conditional mass functions at $z=1.032$ for different parent masses (as shown above each panel) at $z=0$. Filled green circles indicate results from the \protect\MDPL\ simulation, blue open squares indicate results from the maximum posterior likelihood model found in this work, and yellow triangles indicate results obtained using the original parameters proposed by PCH. Small symbols are used in the region where each model is affected by numerical resolution. The vertical grey line indicates the resolution limit of the simulation---points at lower mass ratios are excluded from our fitting procedure, as are points at mass ratios above the peak (if any) of the conditional mass function. Results using the PCH algorithm (including both those generated using the original PCH parameters and the parameters found in this work) are shown convolved with the expected error distributions for halo masses for both parent and progenitor halos.}
 \label{fig:cmfMassSequence}
\end{figure*}

Figure~\ref{fig:cmfMassSequence} shows similar results but now at a fixed redshift of $z=1.032$ and with the parent halo mass varying, as indicated above each panel. Again, it is clear that the PCH algorithm captures the trends in mass seen in the N-body data. 

Performing a \PPC\ for the PCH algorithm we again find that the Bayesian $p$-value is very low ($<10^{-3}$), indicating that this model is formally not a good description of the data.

\section{Discussion}\label{sec:discussion}

We have measured halo mass functions, and conditional mass functions from the \MDPL\ N-body simulation. In contrast to traditional measures of the halo mass function, we explicitly exclude halos which have previously been subhalos inside a larger halo (``backsplash'' halos)---both in the halo mass function and as both parent and progenitor halos in the conditional mass function. Such halos are likely to have different physical characteristics than halos which have always been isolated, since they may have experienced substantial tidal interactions. Furthermore, when halo mass functions coupled with merger tree construction algorithms calibrated to conditional mass functions are used in semi-analytic models to construct the galaxy population, not excluding these backsplash halos leads to a double counting (i.e. they are counted both in the halo mass function, and as progenitors in merger trees of higher mass systems)\footnote{There is an additional subtlety in this case. In merger trees built via extended Press-Schechter-type algorithms, there is generally no mechanism to determine if a given subhalo should be considered a backsplash subhalo at any given time. As such, the masses of backsplash halos are included in the mass of their host halo. This leads to a bias in that halos built in this way are overmassive relative to if backsplash halos were identified and removed. This problem could be circumvented if these extended Press-Schechter merger trees were augmented with a model for the orbital evolution of subhalos---this would allow backsplash subhalos to be identified and their mass removed from their host halo. We expect to address this issue in a future work.}.

While the difference between the $z=0$ halo mass function with backsplash halos included and excluded is only of the level of 10\% over the range of masses probed here, extrapolating to smaller masses suggests that this difference will become larger. Even this 10\% level difference is statistically very significant given the precision with which halo mass functions can now be measured (e.g. the amplitude of the \cite{sheth_excursion_2002} fitting function is determined to a precision of 0.06\%), and is significant compared to the precision of observational data to which semi-analytic models based on PCH merger trees are being constained \citep{bower_parameter_2010,lu_bayesian_2012,lu_bayesian_2014,benson_building_2014}. Of course, the problem of how to define a halo remains---as illustrated in Fig.~\ref{fig:massFunctions} the choice of halo finder can affect the mass function at the level of a few percent (see also \citealt{knebe_structure_2013}).

Furthermore, when fitting functional forms to the mass function and conditional mass functions we convolve with the expected mass-dependent error distribution of N-body dark matter halo masses which arises from the particle nature of N-body simulations. As shown in Fig.~\ref{fig:massFunctionConstraints}, given the precision with which N-body halo mass functions can now be measured, ignoring these errors leads to a significant bias in the parameters of halo mass function fitting functions. Importantly, since the error at given halo mass depends on the number of particles in such halos, ignoring these errors would also make the best fit parameter values dependent on the resolution of the N-body simulation. The \MDPL\ uses cosmological parameters consistent with measurements from the \emph{Planck} satellite \citep{planck_collaboration_planck_2014}, but since the fitting functions we employ make use of scale-free quantities (e.g. $\nu$ as defined in \S\ref{sec:Introduction}) we expect them to be mostly independent of the choice of cosmological parameters (e.g. \citealt{despali_universality_2015}).

To obtain calibrated halo merger rates consistent with the halo mass function, we utilize the algorithm of PCH to generate conditional mass functions of halos selected at $z=0$ and constrain the parameters of this algorithm to match conditional mass functions measured from the \MDPL\ across a range of parent halo masses and progenitor halo redshifts. We obtain tight constraints on the parameters of this model, which differ significantly from previous estimates due to the lack of treatment of N-body halo mass errors in previous work, together with differences in our choice of goodness-of-fit metric, and the different cosmology of our simulation (although this latter effect is expected to be weak due to the way the PCH algorithm is parameterized). While the parameters are formally well-constrained, the PCH algorithm displays notable failings in its ability to match the results of N-body simulations. In particular, it underpredicts the number of high-mass progenitors, and fails to match the shape of the conditional mass function for small time intervals.

The PCH modification to merger rates can be viewed as the first terms in a Taylor series expansion of a general modifier function \citep{parkinson_generating_2008}. Therefore, in principle we could add additional terms (or simply find an entirely different functional form) and presumably find a better match to the data. However, this would complicate the algorithm numerically (and likely reduce its speed substantially). Given that N-body conditional mass functions may themselves be unreliable \citep{srisawat_sussing_2013,jiang_generating_2014} in the regimes where they currently disagree with the PCH algorithm we do not advocate for such additional complexity to be introduced until the N-body results are more robust. In particular, a method to reliably identify backsplash halos would be very advantageous---the current limitation to such identification is the difficulty in assigning the status of ``main'' branch to halos in merger trees with the majority of halo finding algorithms showing fluctuations in this assignment between halos as successive snapshots (see \S5.3 of \citealt{srisawat_sussing_2013}).

\section{Conclusions}\label{sec:conclusions}

Given the levels of precision now achievable in N-body simulations it becomes important for models of structure formation based on Monte Carlo merger trees to use consistently-derived calibrations of merger tree branching probability rates and halo mass functions. The results presented in this work provide such a consistent pair of calibrations for the {\sc RockStar} halo finder and {\sc ConsistentTrees} tree builder applied to a simulation utilizing up to date cosmological parameters. Specifically, our calibrations are to mass functions from which backsplash halos have been excluded, thereby avoiding the double-counting of backsplash halo mass which otherwise occurs. Furthermore, we derive a simple expression for the expected error distribution of N-body spherical overdensity halo masses arising from the particle nature of N-body simulations, and convolve fitting functions with this distribution when constraining their parameters. We show that this results in a significant shift in the best-fit values of mass function parameters, even when restricting the fit to well resolved ($N>300$ particle) halos.

While neither the \cite{sheth_excursion_2002} fitting function, nor the PCH algorithm employed here are formally good-fits to the relevant N-body data as judged by posterior predictive checks, both are adequate fits in the sense that they are usefully close to reproducing the qualitative and quantitative behaviour seen in the N-body simulations. This poor goodness-of-fit could, of course, be improved by adopting more complex (or, perhaps, just different) models and fitting functions. While that may be necessary it is arguably first necessary to understand the N-body data itself more carefully\footnote{A somewhat less satisfactory, but perhaps more practical, solution would be to assess the level of systematic uncertainty in N-body estimates of mass functions and conditional mass functions and include these into the likelihood function used in constraining models. This would inflate the uncertainty on model parameters, but may show that these models are good fits to the data given the systematic uncertainties.}---the process of inferring merger trees and rates from N-body simulations is by no means a solved problem \citep{knebe_structure_2013,srisawat_sussing_2013,jiang_generating_2014}.

In Benson (in preparation) we have examined the properties of galaxies formed by the \glc\ model in merger trees generated via the PCH algorithm, and in merger trees extracted from N-body simulations. When all other tree properties are matched (i.e. cosmological parameters, mass resolution, temporal resolution, and distribution of halo masses at $z=0$) the predicted galaxy properties still differ significantly. These differences are found to be substantially reduced if the N-body merger trees are forced to be always monotonically growing in mass along each branch. In a future work we will explore whether the N-body halo mass error distribution derived in this work can be applied to PCH trees to result in a closer match to the results of N-body simulations. If so, this would provide both motivation to utilize PCH merger trees for calculations requiring high precision, and a means by which to quantify the bias introduced into results by the noisy nature of N-body halo masses.

\section*{Acknowledgments}

We thank Christoph Behrens for providing the \MDPL\ merger trees in the \glc\ merger tree file format, and Shaun Cole and Yu Lu for helpful discussions. The author gratefully acknowledges the Gauss Centre for Supercomputing e.V. (\href{www.gauss-centre.eu}{\tt www.gauss-centre.eu}) and the Partnership for Advanced Supercomputing in Europe (PRACE, \href{www.prace-ri.eu}{\tt www.prace-ri.eu}) for funding the MultiDark simulation project by providing computing time on the GCS Supercomputer SuperMUC at Leibniz Supercomputing Centre (LRZ, \href{www.lrz.de}{\tt www.lrz.de}). The CosmoSim database used in this paper is a service by the Leibniz-Institute for Astrophysics Potsdam (AIP). The MultiDark database was developed in cooperation with the Spanish MultiDark Consolider Project CSD2009-00064.

\bibliographystyle{mn2e}
\bibliography{treeMassFunctionAccented}

\end{document}